\documentstyle[11pt,amssym,aaspp4]{article}
\received{}
\accepted{November 1998}
\journalid{516}{May 1, 1999}
\articleid{}{}
\cpright{AAS}{1999}
\slugcomment{ApJ, Vol. 516, May 1, 1999}
\lefthead{Ulvestad, Wrobel, \& Carilli}
\righthead{Radio Outflow and Absorption in Seyfert Mrk~231}
\def\H2{\ion{H}{2}}

\begin{document}

\title{Radio Continuum Evidence \\
       for Outflow and Absorption \\
       in the Seyfert~1 Galaxy Markarian~231}

\author{J.S.~Ulvestad, J.M.~Wrobel, \& C.L.~Carilli}
\affil{National Radio Astronomy Observatory\footnote{The National
Radio Astronomy Observatory is a facility of the National
Science Foundation operated under cooperative agreement
by Associated Universities, Inc.} \\
P.O. Box O, Socorro, NM 87801; \\
julvesta,jwrobel,ccarilli@nrao.edu}

\begin{abstract}
The Very Long Baseline Array (VLBA) and the Very Large Array 
(VLA) have been used to image the continuum radio emission from
Mrk~231, a Seyfert~1 galaxy and the brightest infrared galaxy in the
local universe.  The smallest VLBA scales reveal a double, or possibly
triple, source less than 2~pc in extent.  The components of this
central source have minimum brightness temperatures of
$10^9$--$10^{10}$~K, spectral turnovers between 2 and 10~GHz, and
appear to define the galaxy nucleus plus the inner regions of a jet.
The strongest component is probably synchrotron self-absorbed, while
the weaker component to the northeast may be either free-free absorbed
or synchrotron self-absorbed.  

On larger VLBA scales, the images
confirm a previously known north-south 
triple source extending 40~pc and elongated
perpendicular to a 350-pc starburst disk traced by H{\sc i} and CO.
Both lobes of the triple show evidence for free-free absorption near
2~GHz, probably due to ionized gas with a density of 1--2$\times
10^3$~cm$^{-3}$ in the innermost regions of the starburst disk.  This
free-free absorption resembles that toward the counterjet of 3C~84,
but requires ionized gas at lower density located considerably farther
from the central source.  The absorbing gas may be ionized by the
active nucleus or by local regions of enhanced star formation,
possibly in the inner part of the starburst disk.
  The elongation position angle of the 40-pc
triple differs by 65\arcdeg\ from that of the 2-pc source.
Unless the radio source is seen nearly end-on, 
the different symmetry axes on different scales in Mrk~231 imply a dramatic
curvature in the inner part of the Mrk~231 radio jet.

A comparison of VLBA and VLA flux densities indicates that 
the radio continuum from the 350-pc disk has a spectral
index near $-0.4$ at frequencies
above 1.4~GHz and is plausibly energized by a
massive burst of star formation, with the overall spectrum
flattened somewhat by a contribution from free-free absorption.  
On VLA scales, asymmetric and diffuse
emission extends for more than 25~kpc.  This emission has a steep
spectrum, exhibits linear polarization exceeding 50\% at some
locations, and shares the symmetry axis of the 40-pc triple, but on a scale
larger by three orders of magnitude.  The large-scale radio emission 
extends beyond the bulk of the optical galaxy, but has an initial axis similar to a
series of optical star-forming knots several kiloparsecs from
the nucleus.  This diffuse radio source is probably
generated by energy deposition from a slow-moving nuclear jet, 
which conceivably could help energize the off-nuclear starburst as well.
\end{abstract}


\section{Introduction}

Classified optically as a Seyfert~1 galaxy, Markarian~231 is also the
most luminous infrared galaxy in the local ($z<0.1$) universe
(\cite{sur98}).  Ultraluminous infrared galaxies like Mrk~231 have
total infrared luminosities well above $10^{11}~L_\odot$, as measured
by IRAS.  Such galaxies are thought to be stages along a sequence in
the evolution of merging spiral galaxies; the mergers generate
enormous bursts of star formation and the merged galaxies eventually
turn into quasars (e.g., \cite{san88}).  Both the enormous infrared
luminosity of $\sim 3\times 10^{12}~L_\odot$ (\cite{soi89};
\cite{bon97}) and the Seyfert/quasar properties of Mrk~231 argue that
it is well along in the merger sequence.  Using a velocity relative to
the 3~K background of 12,447~km~s$^{-1}$ (\cite{dev91}) and $H_0 =
75$~km~s$^{-1}$~Mpc$^{-1}$, the distance to Mrk~231 is 166~Mpc, and
800~pc subtend 1\arcsec.

On a galactic scale, Mrk~231 shows tidal tails indicative of a recent
interaction (e.g., \cite{hut87}; \cite{lip94}; \cite{sur98}).  Optical
emission lines from
an apparent star-forming region are seen 10\arcsec--15\arcsec\ 
($\sim 10$~kpc) south of
the nucleus (\cite{ham87}; \cite{hut87}).  An extended radio continuum
source roughly 60\arcsec\ (48~kpc) in diameter is also present,
predominantly to the south of the nucleus (\cite{deb76}; \cite{hut87};
\cite{con88}).  Armus et al.\ (1994) reported a second
nucleus about 3.5\arcsec\ south of the main nucleus, but 
{\it Hubble Space Telescope\/} (HST) imaging
shows only a series of star-forming knots at this location
(\cite{sur98}).  There is a general consensus that many of the
observed properties of the galaxy result from a merger occurring
$10^8$--$10^9$~years ago (e.g., \cite{hut87}; \cite{arm94};
\cite{sur98}).  Baum et al.\ (1993) imaged a
continuum source at 1.4~GHz that extends for 150\arcsec\ (120~kpc)
perpendicular to the galaxy disk, and interpreted that source as a
superwind driven by the intense star formation triggered by the
merger.

The inner region of Mrk~231 contains a CO disk aligned almost
east-west, with an inner diameter of about 1\arcsec\ (800~pc),
a lower density region extending to 3\arcsec, and a total gas
mass exceeding $10^9~M_\odot$ (\cite{bry96}; \cite{dow98}).  
A compact 10-$\mu$m
source has a maximum diameter of 0.6\arcsec\ (\cite{mil96}) and is
apparently associated with the active galactic nucleus (AGN).  
That nucleus shows
broad Balmer emission lines, characteristic of a Seyfert~1 nucleus,
rather weak forbidden lines (e.g., \cite{bok77}), and variable,
low-ionization, and broad absorption lines at redshifts up to
7800~km~s$^{-1}$ relative to the systemic velocity (\cite{bor91};
\cite{for95}).  The nucleus
shows strong Fe~II emission, although its X-ray and low-energy
$\gamma$-ray emission are anomalously weak for an AGN
(\cite{rig96}; \cite{der97}; \cite{law97}).  ASCA
observations imply a ratio $L_x/L_{\rm FIR} = 7\times 10^{-4}$, more
consistent with a starburst galaxy than with an AGN,
although there clearly is a hard power-law component associated
with the active nucleus (\cite{nak97}).  The ASCA data show an X-ray
absorbing column of $N_H=6\times 10^{22}$~cm$^{-2}$, which may 
be associated with the gas in the broad-line region.  OH maser emission
with an isotropic luminosity of $700~L_\odot$ also is associated with the
galaxy (\cite{baa85}) but no H$_2$O megamaser emission has been detected
from the nucleus, with an upper limit of $18~L_\odot$ (\cite{bra96}).
The OH emission is spread over a scale of a few hundred parsecs,
rather than being confined to a very compact region as in most
other active galaxies (\cite{lon98}).

Numerous Very Large Array (VLA) images of Mrk~231 are 
available at $\sim$1\arcsec\ or
better resolution (e.g., \cite{ulv81}; \cite{nef88}; \cite{con91}; \cite{pat92};
\cite{kuk95}; \cite{pap95}).  These images are dominated by an unresolved 
VLA core with
a typical flux density of 100--300~mJy between 1 and 22~GHz; based
on the available published data, this core appears to vary at 
gigahertz frequencies by tens of percent on time scales of years. 
There also is very weak extended emission on the sub-arcsecond scale
(\cite{tay99}).  The H{\sc i}
absorption against the VLA core (\cite{dic82}) occurs
against a radio halo or disk of emission 440~mas (350~pc) in extent, with an
elongation PA in the east-west direction (Carilli, Wrobel, \&
Ulvestad 1998, hereafter CWU), similar to
the elongation PA of the 800-mas CO emission (\cite{bry96}).  The
radio continuum from this 350-pc disk is strongest at low frequencies
and is apparently responsible for much of the extra $\sim 100$~mJy of flux
density missed between measurements on VLBI scales and those on VLA
scales (e.g., \cite{lon93}; \cite{tay94}; \cite{tay99}).  The properties of the
350-pc disk imply a massive star formation rate of
$60~M_\odot$~yr$^{-1}$ (\cite{car98}).  On a smaller
scale of 50~mas (40~pc), the core resolves into a north-south triple at
1.7~GHz, as imaged with the European VLBI Network in the mid-1980s
(\cite{nef88}).  Optical and ultraviolet spectropolarimetry give an
electric-vector position angle (PA) of $\sim 95$\arcdeg, suggesting
scattering of radiation by dust clouds flowing outward along
this north-south VLBI axis (\cite{goo94}; \cite{smi95}).  

This paper presents and interprets new continuum observations, with
the Very Long Baseline Array (VLBA) and the VLA, 
designed to probe structures in Mrk~231 on scales
ranging from parsecs to kiloparsecs.  Preliminary results were
reported by Ulvestad, Wrobel, \& Carilli (1998).  This radio continuum
study of Mrk~231 provides evidence for outflow, for synchrotron
self-absorption, and for free-free absorption.  These new radio
results are related, wherever possible, to published results from the
aforementioned studies of radio spectral lines and in shorter wavelength bands.

\section{Observations and Calibration}

\subsection{VLBA}
\label{vlbaobs}

The continuum emission from Mrk~231 was observed with the 10-element VLBA
(\cite{nap94}) at frequencies ranging from 1.4 to 22.2~GHz, 
during the three separate observing sessions
summarized in Table~\ref{tab:vlbaobs}.  During 1995 November,
11-minute scans on Mrk~231 typically were preceded by 2-minute
scans on the delay-rate check source J1219+4829, with a total time of
about 39~minutes to cycle through 1.4, 2.3, and 5.0~GHz.  Occasional
scans on DA~193 (J0555+3948), 3C~345 (J1642+3949), and J1740+5212 were
obtained for ancillary calibration (fringe-finding, manual
pulse calibration, and amplitude calibration checks).  On 1996
December 8, scans of 12~minutes duration on Mrk~231 were interleaved
with 2--4-minute scans on J1219+4829, and the total cycle time for 5.0,
8.4, and 15~GHz was 44~minutes.  A single 25-m VLA antenna was included
along with the VLBA to provide improved short-spacing coverage.  
Ancillary calibrators were 4C~39.25
(J0927+3902), J1310+3220, and 3C~345.  On 1996 December 27, both
H{\sc i} line and 22-GHz continuum data were acquired using the
VLBA and the phased VLA.  The line
observations were reported by CWU and
will not be described further here.  At 22 GHz, 8-minute scans of
Mrk~231 were interleaved with 3-minute scans of J1219+4829.  Short
scans of 4C~39.25, J1310+3220, and 3C~345 were included for ancillary
calibration.  All VLBA observing and correlation of Mrk~231 adopted
the J2000 position for source J1256+5652 from Patnaik et al.\ (1992).
At the VLA, additional scans of 3C\,286 (J1331+3030) were obtained 
on 1996 December 27 to
set the flux density scale, and scans of Mrk~231 at 5, 8.4, and 15~GHz
were also acquired to provide a contemporaneous spectrum of the
galaxy.

Initial calibration of the VLBA data was carried out using the
standard gain values of the VLBA antennas together with system
temperatures measured every 1--2 minutes during the observing
sessions.  Autocorrelations were used to correct for imperfect
adjustment of the sampler levels for these 2-bit (4-level) data.  For
the 1995 data, an additional amplitude adjustment was made using the
VLBA image of DA~193, whose total flux density was constrained to be
equal to the value measured at the VLA.  Baseline gain corrections
were less than 5\% at 5~GHz, 5--10\% at 2.3~GHz, and 5--20\% at
1.4~GHz.  The substantial corrections at 1.4~GHz were expected because
the observing frequency differed by nearly 300~MHz (roughly 20\% of
the observing frequency) from the frequency where standard gains are
measured.  Since the VLBA gains are quite stable, and accurate system
temperatures were regularly measured, the estimated uncertainty in the
VLBA flux density scale is less than $\sim 10$\% .  The NRAO
Astronomical Image Processing System (AIPS) (\cite{van96}) was used
for all VLBA data calibration.

\subsection{VLA}
\label{vlaobs}

The continuum emission from Mrk~231 was observed in dual circular
polarizations with the VLA (\cite{tho80}) during the five separate
observing sessions summarized in Table~\ref{tab:vlaobs}.  In 1988 and
1989, scaled-array observations were made at 1.5, 4.8, and 15.0~GHz,
using the B, C, and D configurations, respectively.  These
observations yielded intrinsic resolution near 4\arcsec, useful for
studying the emission on kiloparsec scales.  Scans of $\sim
30$~minutes were interleaved with scans of J1219+4829, the local phase
calibrator, and the amplitude scale was set to that of Baars et al.\
(1977) using short observations of 3C~286.  These scaled-array
observations were not full syntheses, as only 1.5--3~hr were spent on
Mrk~231 in each case.  During 1995 November, a short observation in the
B configuration was made to determine the total flux and spectrum of
Mrk~231 contemporaneously with the 1995 VLBA observations.  Five
frequencies between 1.4 and 22~GHz were used at the VLA, with 11
minutes spent on Mrk~231 at each frequency.  Resolutions ranged from
$\sim 4$\arcsec\ at 1.5 GHz to $\sim 0.3$\arcsec\ at 22~GHz.
J1219+4829 was observed for several minutes at each frequency as a
phase calibrator.  Also, short scans of 3C~286 were obtained to set
the VLA flux density scale and short scans of DA~193 were acquired to
check the VLBA flux density scale.  During 1996 December, observations
of Mrk~231 at 5, 8.4, and 15~GHz were only single 2-minute snapshots,
while the 1.4- and 22-GHz observations made together with the
VLBA were much longer.  Phase and amplitude calibration using
J1219+4829 and 3C~286, respectively, was similar to that carried out
for earlier VLA epochs, with the added complication that additional
editing was required for calibrator scans acquired in phased-array  
mode.  VLA flux densities typically have errors of $\sim 5$\% but
residual phase noise in 1996 led to larger errors at the higher
frequencies (7.5\% at 8.4~GHz, 10\% at 15~GHz, and 15\% at 22~GHz).
At 1.4 GHz, a polarization calibration was performed using J1219+4829
to deduce the antenna polarizations and using 3C~286 to fix the
absolute polarization PA, and the 1.4-GHz data were corrected for
these effects.  AIPS was used for all VLA data calibration.

\section{VLBA Imaging}

Following the initial calibration, all VLBA data were imaged in AIPS.
After initial images were made, an iterative self-calibration
procedure was used to correct the complex gains of the individual
antennas for atmospheric and instrumental effects.  This process
halted when the image quality and the r.m.s. noise stabilized.

A major aim of this study was to derive the spectra of different
components in Mrk~231 to search for absorption.  To accomplish this
goal, full-resolution images were made at each VLBA frequency from 1.4
to 15~GHz.  Then, the data at each frequency were re-imaged using only
the range of projected baseline lengths sampled at each of the lower
frequencies.  In this process, the weighting of the data was tapered
in the aperture plane to give approximately the same resolution as
that available at each lower frequency, with a restoring beam fixed to
have the same parameters as the full resolution beam at the lower
frequency.  For example, the 8.4-GHz data were re-imaged at three
different resolutions, equivalent to the full resolution at 1.4, 2.3,
and 5.0~GHz.  The subsections below present a subset of images that
are the most important for the scientific analysis.  The
interpretation of the results is deferred to later sections.

\subsection{North-South Triple}

\begin{figure}[th]
\vspace{18.5cm}
\includegraphics{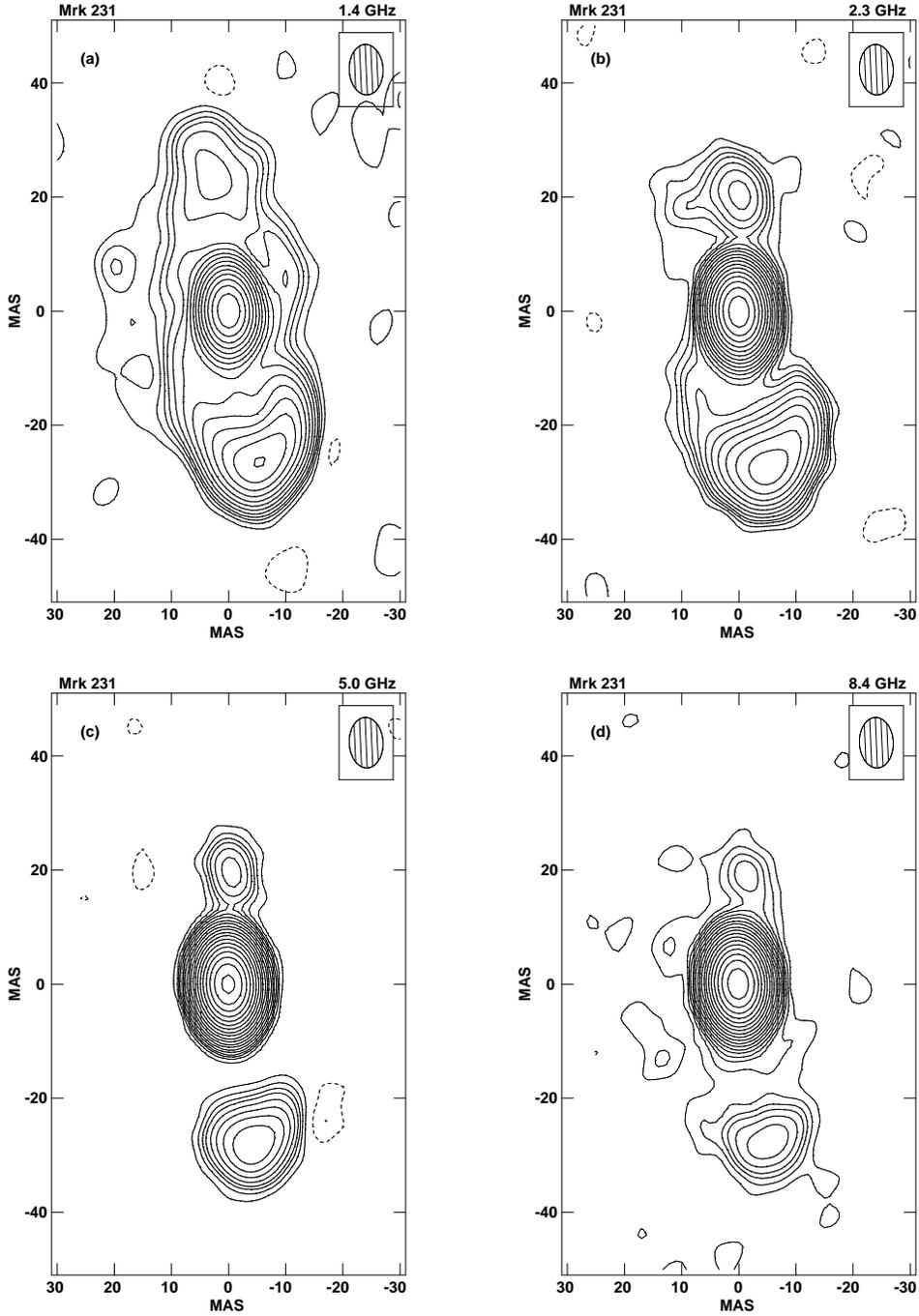}
\caption{CLEANed VLBA images at matched resolution, all shown
at the same scale.  In these
and all succeeding images, North is up and East is to the left, with
negative contours shown as dashed lines.  Hatched region
shows the elliptical Gaussian restoring beam at FWHM of $8.85\times
5.92$~mas elongated in PA 2.6\arcdeg.  Contour levels are plotted 
at 200~$\mu$Jy~beam$^{-1}$ times $-4$, $-3$, $-2$, $-1$, 1, 2, 3, and 4, with
higher levels spaced by factors of $2^{1/2}$.  
(a) Epoch 1995. Frequency 1.4~GHz. Peak intensity  48.8~mJy~beam$^{-1}$.
(b) Epoch 1995. Frequency 2.3~GHz. Peak intensity  93.9~mJy~beam$^{-1}$.
(c) Epoch 1995. Frequency 5.0~GHz. Peak intensity 160.7~mJy~beam$^{-1}$.
(d) Epoch 1996. Frequency 8.4~GHz. Peak intensity 133.0~mJy~beam$^{-1}$.
\label{fig:vmatch}}
\end{figure}

The new VLBA image at 1.4 GHz, presented in Figure~\ref{fig:vmatch},
shows the same (nearly) north-south triple structure known from
previous 1.7-GHz observations with the European VLBI Network
(\cite{nef88}).  The triple consists of an unresolved core, together
with two resolved lobes.  The approximate size of the triple is 50~mas
(40~pc).  The emission 20~mas to the east of the core is not apparent
at other frequencies, and is likely to be an artifact.  The other
images in Figure~\ref{fig:vmatch} are the 2.3, 5.0, and
8.4-GHz images of the same triple at a resolution matching that of the
1.4-GHz image.  Note that as the frequency increases, progressively
less diffuse emission is seen; only the outer portion of the VLBA
lobes is detected at 8.4~GHz.

\begin{figure}[th]
\vspace{18.5cm}
\includegraphics{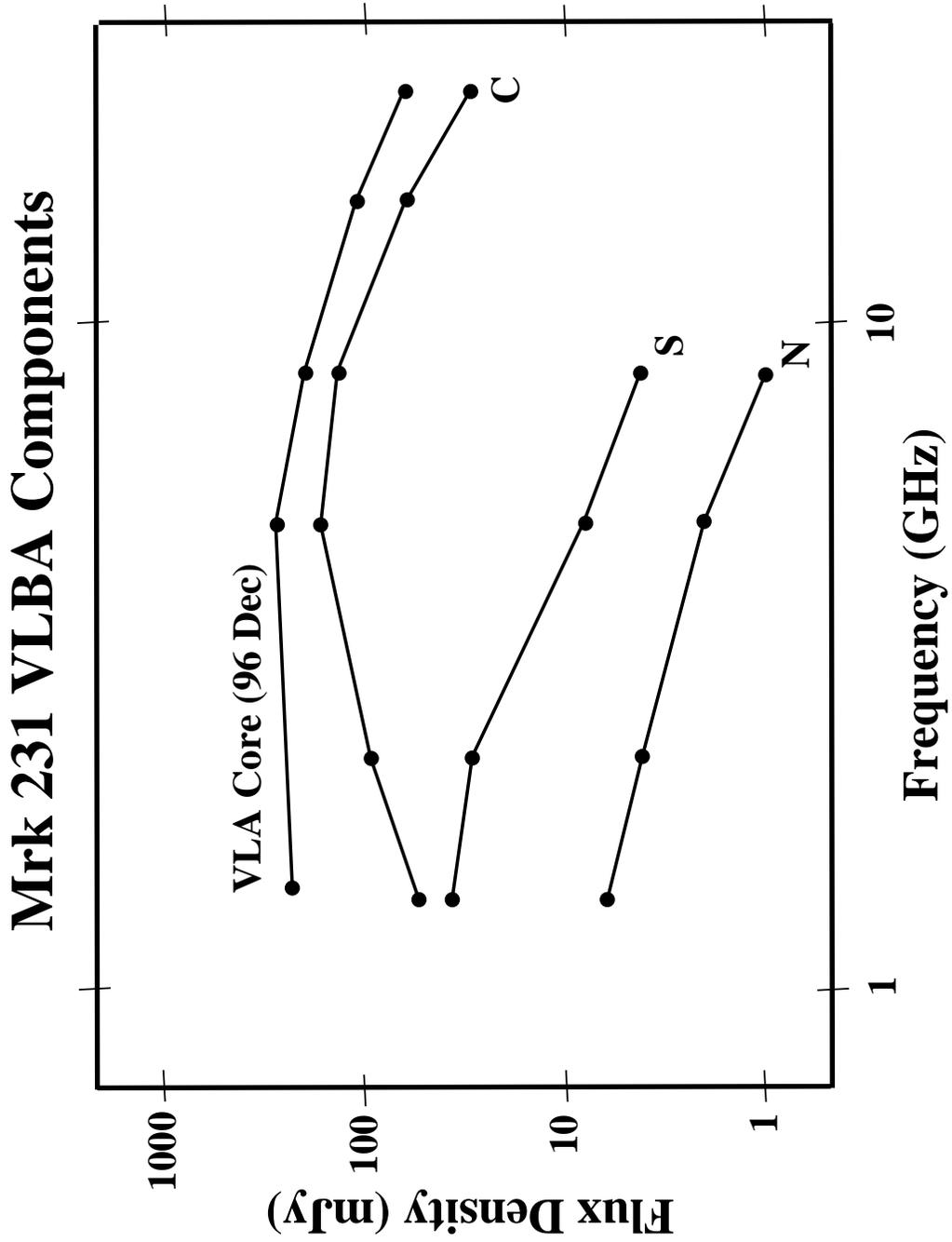}
\caption{Spectra at matched resolution of the three VLBA
components in Figure~\ref{fig:vmatch}, with the VLA core
flux density added for comparison.
\label{fig:vspectra}}
\end{figure}

The total flux densities in the north (N), central (C), and south (S)
components were measured by integrating over the entire area of each
component.  In addition, the position of the peak intensity in each
component, relative to component C, was determined by fitting a
parabola to a few pixels in each image surrounding the peak.  Results
of these measurements are given in Table~\ref{tab:nsflux}.  Also
included in Table~\ref{tab:nsflux} are the results for component C
from the tapered data at 15~GHz; components N and S are too weak to
be detected at that frequency.  The estimated errors in the flux
densities are 10\% for component C and for the low-frequency data on
component S, but rise to as much as 50\% for components N and S at
8.4~GHz due to the increasing uncertainty in the lobe strengths at
frequencies where they are largely resolved out.  Spectra of
components N, C, and S are shown in Figure~\ref{fig:vspectra}.
Although the data at all frequencies were not taken simultaneously,
the well-resolved components, N and S, 
are unlikely to have varied over the 13 months
between sessions.  Component C appears to have been roughly constant
from 1.4 to 5~GHz, based on (1) the VLA core results (see
Section~\ref{yflux}); and (2) the small differences of only 
5--10~mJy between 1995 and 1996 in the VLBA strengths at 1.4 and 5~GHz
(Table~\ref{tab:nsflux}; \cite{car98}).
The data between 5 and 22~GHz were taken within 20~days, so
Figure~\ref{fig:vspectra} should be a good snapshot of the spectrum of
the VLBA components at a single epoch, 1996 December.

\begin{figure}[th]
\vspace{16.0cm}
\includegraphics{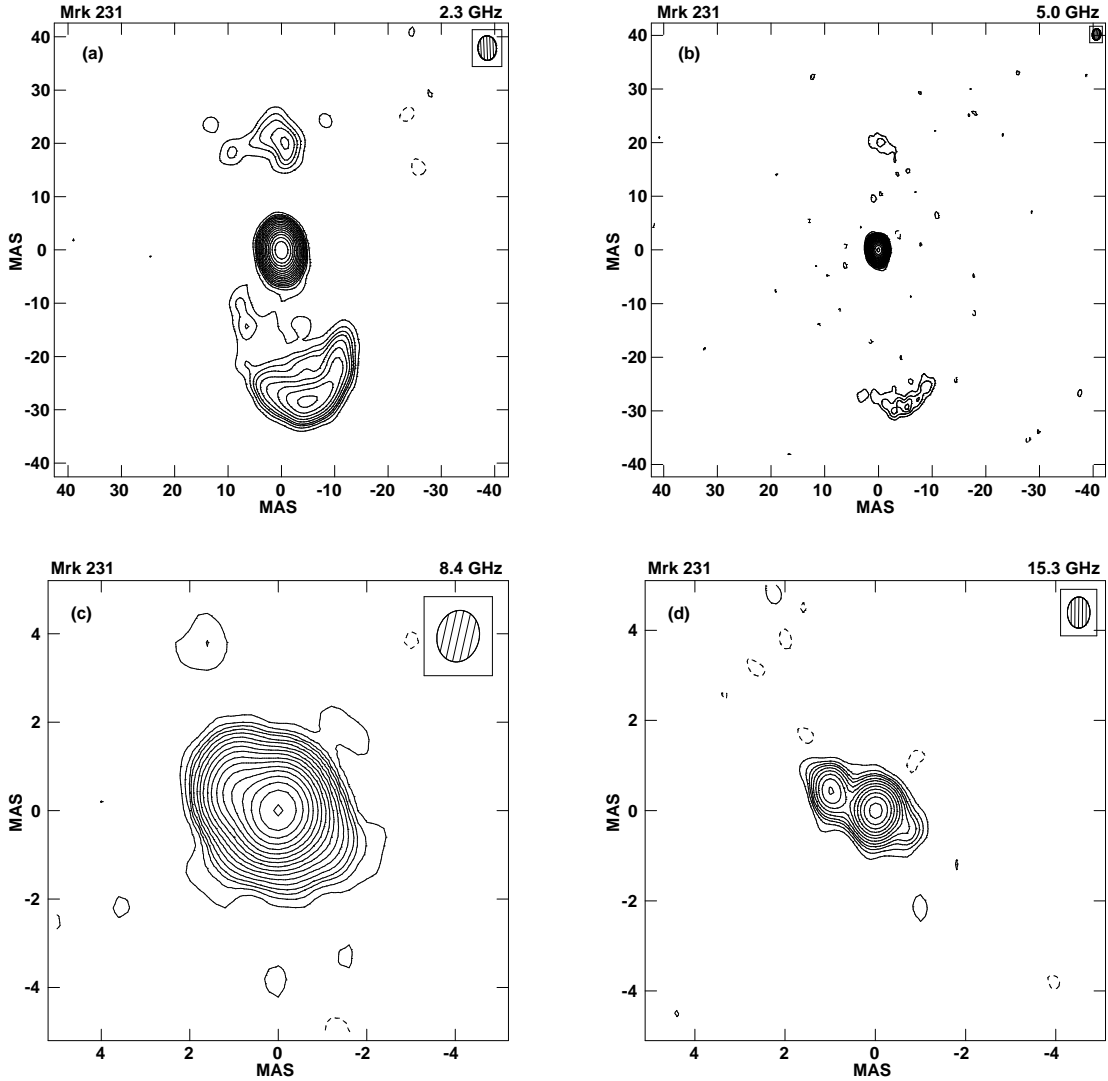}
\caption{CLEANed VLBA images at full resolution.  Hatched regions
show the elliptical Gaussian restoring beams at FWHM.  Contour levels
are the same multiples of the lowest contour as in 
Figure~\ref{fig:vmatch}.  Panels (a) and (b) are shown at the same scale.
Panels (c) and (d) are magnifications of the central component, C,
shown at a common scale.
(a) Epoch 1995. Frequency 2.3~GHz. Peak intensity 89.5~mJy~beam$^{-1}$.
    Lowest positive contour 250~$\mu$Jy~beam$^{-1}$.
    Restoring beam is $4.62\times 3.46$~mas in PA 2.2\arcdeg.  
(b) Epoch 1995. Frequency 5.0~GHz. Peak intensity 140.0~mJy~beam$^{-1}$.
    Lowest positive contour 250~$\mu$Jy~beam$^{-1}$.
    Restoring beam is $2.08\times 1.47$~mas in PA $-5.1$\arcdeg.
(c) Epoch 1996. Frequency 8.4~GHz. Peak intensity 94.9~mJy~beam$^{-1}$.
    Lowest positive contour 250~$\mu$Jy~beam$^{-1}$.
    Restoring beam is $1.17\times 0.95$~mas in PA $-12.5$\arcdeg.
(d) Epoch 1996. Frequency 15.3~GHz. Peak intensity 37.9~mJy~beam$^{-1}$.
    Lowest positive contour 500~$\mu$Jy~beam$^{-1}$.
    Restoring beam is $0.69\times 0.50$~mas in PA $-0.6$\arcdeg.
\label{fig:vfull}}
\end{figure}

VLBA images, at full resolution, of the north-south triple at 2.3 and
5.0~GHz are shown in Figures~\ref{fig:vfull}a and \ref{fig:vfull}b.  
(The full-resolution
image at 8.4 GHz, discussed below, shows no emission from components N
or S.)  Both of these images show the structure at the ends of the
VLBA lobes more clearly, with substantial resolution perpendicular to
the direction to the central source.  Neither the total component flux
densities nor the peak positions differ significantly from those
values derived from the images at matched resolution.

\subsection{Central Component of Triple}
\label{central}

The first installment of the Caltech-Jodrell VLBI survey presented a
5.0-GHz image of Mrk~231, under the alias 1254+571 (\cite{tay94}).
That survey included intercontinental baselines, and indicated that
Mrk~231 was slightly resolved on a scale near 1~mas, in a nearly
east-west direction, markedly different from the PA of the larger,
north-south triple described above.

New VLBA images, at full resolution, of component C at 8.4 and
15.4~GHz are shown in Figures~\ref{fig:vfull}c and \ref{fig:vfull}d.  
The 8.4-GHz image shows
clear resolution in a PA between 60\arcdeg\ and 65\arcdeg, somewhat
north of the PA of 92\arcdeg\ quoted by Taylor et al.\ (1994) for the
resolved core at 5~GHz, although their image does show an extension slightly
north of east, in PA $\sim 80$\arcdeg.  The
15-GHz image shows that the central source appears to break up into
three separate components at sub-parsec resolution.  Single- and
multi-component Gaussian fits were made to component C at all
frequencies between 1.4 and 15~GHz.  At 1.4 and 2.3~GHz, the source is
not resolved, and the total flux densities are indistinguishable from
the values given in Table~\ref{tab:nsflux}.  However, at 5, 8.4, and
15.4~GHz, component C is significantly resolved with a size of
0.8--1.0~mas (0.6--0.8~pc) in a PA near 63\arcdeg.  
At the 8.4-GHz resolution, component C in both
the 8.4 and 15-GHz images is much better fitted by a two-component
model. Finally, at
the full 15-GHz resolution, there may be a third component to the
southwest.  All the fits to component C at the three different
resolutions are summarized 
in Table~\ref{tab:cflux}.  It is possible that the southwestern
component is an imaging artifact, but since its location and that of
the northeastern component are not symmetric with respect to the
central component, the southwestern component is most likely real.  In
the multiple component fits, only the strongest source appears
significantly resolved.  Since the resolution is predominantly along
the direction of the structure, this indicates the possible presence
of more components that would be separated from the strongest source
at yet higher spatial resolution.

Although Mrk~231 also was observed with the VLBA and the phased VLA at
22.2~GHz, the steepening high-frequency spectrum of the core at higher
frequencies meant that fringes were detected only on the relatively
short baselines in the southwestern United States.  Imaging of these limited
data yields a total flux density of 30~mJy for the core, and that
datum is included in the spectral plot given in
Figure~\ref{fig:vspectra}.  Phase-referencing observations are
required to image component C at 22~GHz with the resolution and
sensitivity needed to identify subcomponents.

\section{VLA Imaging}

\subsection{Scaled Arrays}

\begin{figure}[th]
\vspace{12.5cm}
\includegraphics{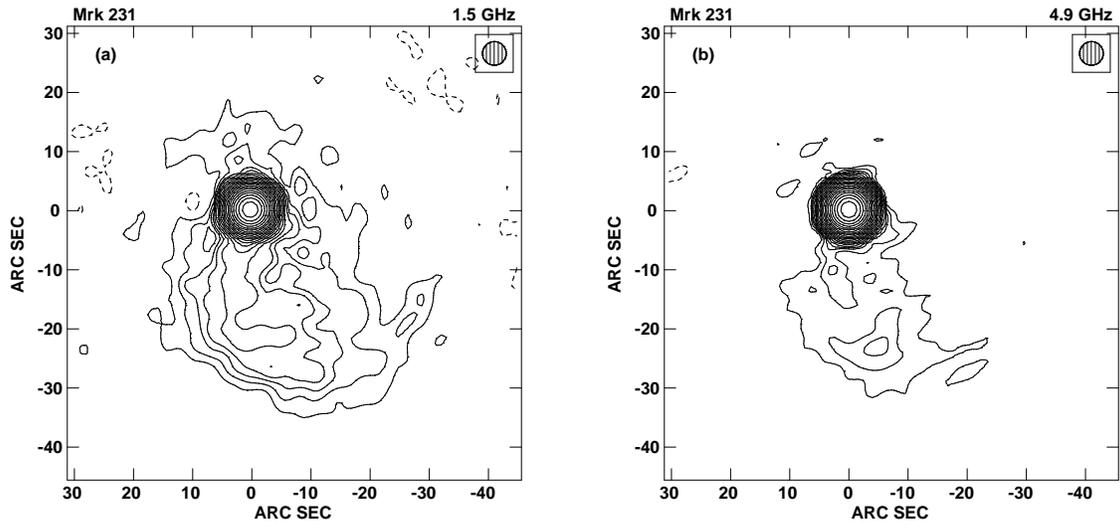}
\caption{CLEANed VLA images of Stokes I, at matched resolution, at
epoch 1989.  Hatched region shows the circular Gaussian restoring beam
at FWHM of 4\arcsec.  Relative contour levels are as in
Figure~\ref{fig:vmatch}, with the lowest positive contour at 
170~$\mu$Jy~beam$^{-1}$.  
(a) Frequency 1.5~GHz. Peak intensity 231.9~mJy~beam$^{-1}$.
(b) Frequency 4.9~GHz. Peak intensity 340.3~mJy~beam$^{-1}$.
\label{fig:ymatch}}
\end{figure}

During 1988 and 1989, Mrk~231 was observed at 1.5, 4.9, and 15~GHz
using scaled arrays of the VLA, as described in Section~\ref{vlaobs}.
Images were restored with a common (circular)
Gaussian beam of size 4\arcsec\ (full width at half maximum), 
to enable spectral comparisons.
Figure~\ref{fig:ymatch} shows these images at 1.5 GHz and 4.9 GHz.
The 15-GHz emission is completely unresolved, so that image is not
shown.  Flux densities in the unresolved VLA core were derived by
fitting a Gaussian constrained to the beam size to the central pixels
of each image.  In addition, the total flux density in each VLA image
was determined by integrating over the region in which
significant emission is detected; the extended flux density, then, is
taken to be the difference between the total flux density and that in
the unresolved VLA core.  At 15~GHz, there is no detection of extended
emission in a single beam area, so this difference is taken to be an
upper limit to the total extended flux density.  Peak intensities in
the extended emission were measured at 1.5 and 4.9 GHz, while an upper limit of
3 times the r.m.s. noise per beam area was assumed at 15~GHz.  Results
of all these flux density measurements appear in
Table~\ref{tab:vlascale}.

The spectral index of the extended emission is steep, both in total
flux density and in a point-by-point comparison.  The two-point
spectral index, $\alpha$ of the total emission is $-1.05\pm 0.23$ between 1.5
and 4.9~GHz ($S_\nu\propto \nu^{+\alpha}$, where $S\nu$ is the flux
density at frequency $\nu$).  A spectral-index image between these frequencies
indicates that all regions with significant emission at both
frequencies have spectral indices ranging between $-0.4$ and $-1.0$,
with the total spectrum being somewhat steeper because of the areas
detected at 1.5~GHz that are below the detection threshold at 4.9~GHz.
The total spectral index of the extended emission between 4.9 and
15~GHz is $\alpha <-0.8\pm 0.4$.

\subsection{Flux Density Monitoring}
\label{yflux}

VLA observations of Mrk~231 were made in 1995 November and
1996~December, at frequencies ranging from 1.4~GHz to 22~GHz, as
described previously.  The data were calibrated, imaged, and
self-calibrated in the usual way, in order to measure the
flux density of the unresolved VLA core.  This flux density was
taken to be the peak 
in the final self-calibrated images.  Results are presented in
Table~\ref{tab:vlaflux}, with errors quoted as discussed in
Section~\ref{vlaobs}.  The measurements show that the flux density of
the core appears to have been relatively constant at frequencies
up to 5~GHz, but decreased significantly at 15 and 22~GHz from 1995 to
1996.  Therefore, the spectrum of the central component of the VLBA
triple source (see Figure~\ref{fig:vspectra}) should be valid as of
the 1996 December epoch.  Weak extended emission in the vicinity of
the core, within the central arcsecond, is present in the highest
resolution VLA data, as discussed by Taylor et al. (1998).  This emission
has no significant impact on the flux-density measurements for
the unresolved core.

\subsection{Deep VLA Polarimetry}

\begin{figure}[th]
\vspace{16.0cm}
\includegraphics{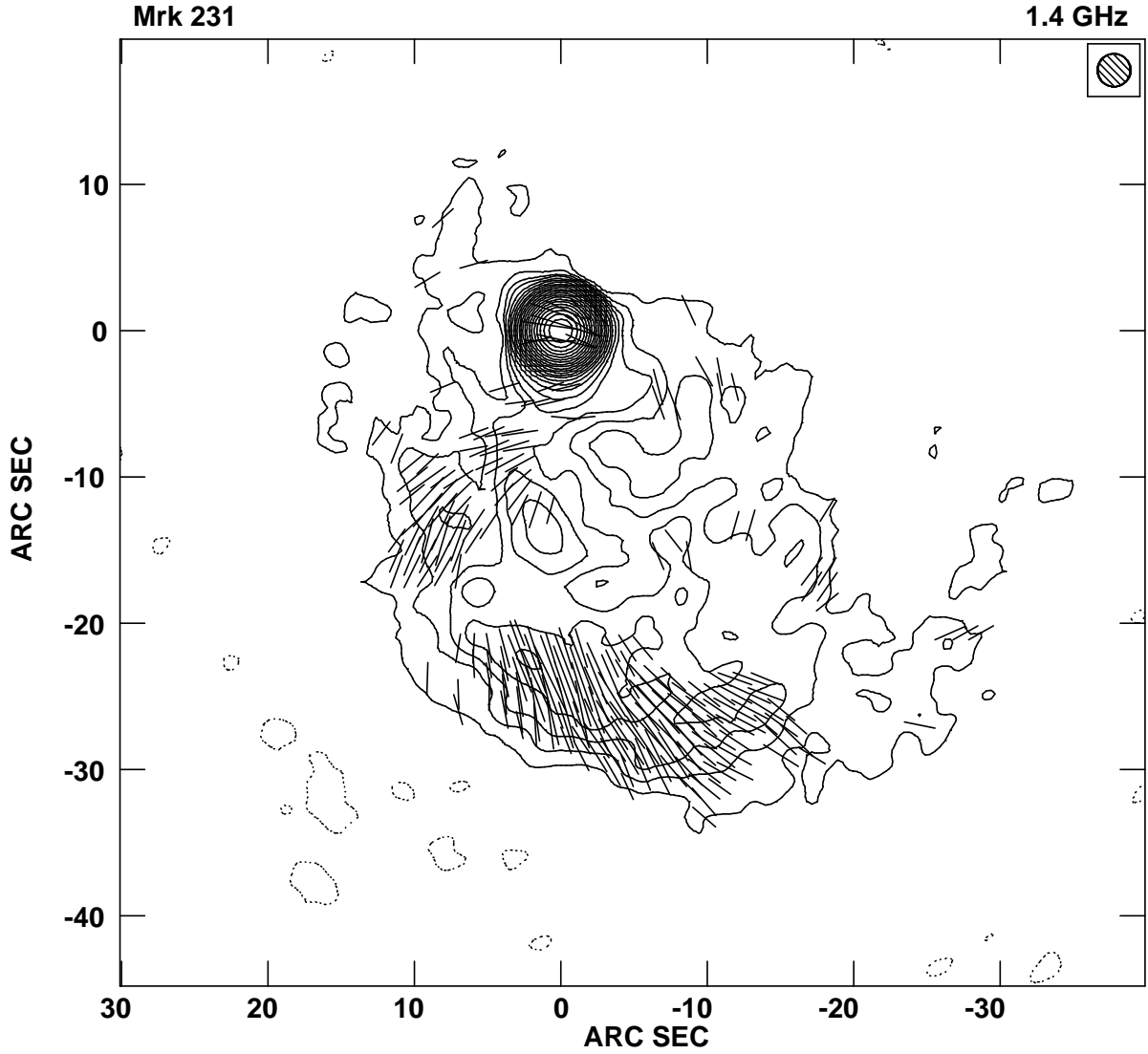}
\caption{CLEANed VLA image of Stokes I at epoch 1996 and frequency
1.4~GHz.  Hatched region shows the circular Gaussian restoring beam at
FWHM of 2.25\arcsec.  Relative contour levels are as in
Figure~\ref{fig:vmatch}.  Lowest positive contour is
80~$\mu$Jy~beam$^{-1}$.  Peak
intensity is 231.4~mJy~beam$^{-1}$.  Line lengths represent linearly
polarized intensity P (1\arcsec\ is 36~$\mu$Jy~beam$^{-1}$).  Line
orientations represent electric-field PA $\chi$.
\label{fig:ydeep}}
\end{figure}

A long observation of Mrk~231 was made at 1.4~GHz with the phased VLA
in 1996 December, as part of a VLBA observation of the H{\sc i}
(CWU).  The phased-array data
essentially undergo a real-time calibration of the VLA phases, and
then can be calibrated further in AIPS, as described previously.
Mrk~231 was imaged in Stokes I, Q, and U; the Q
and U images were further combined in the usual way to obtain images
of the linearly polarized intensity, P, and electric-field PA, $\chi$.
Figure~\ref{fig:ydeep} shows a composite image, with lines
representing P and $\chi$ superposed on contours of Stokes I emission.
The Stokes I emission in Figure~\ref{fig:ydeep} strongly resembles
that evident in Fig.~\ref{fig:ymatch}.  However, a major new discovery
from Figure~\ref{fig:ydeep} is that some regions of the diffuse Stokes
I emission to the south of the VLA core are significantly linearly
polarized, reaching a peak polarized intensity of
185~$\mu$Jy~beam$^{-1}$ about 26\arcsec\ (21~kpc) south of the VLA
core.  At this polarization peak, the percentage polarization is about
57\% and $\chi$ (electric vector position angle) 
is about 15\arcdeg.  No rotation measure corrections
have been made.  The VLA core is $<$ 0.1\% linearly polarized.

\section{Interpretation, from Large to Small Scales}

\subsection{Summary of Radio/Optical Structures}

\begin{figure}[htp]
\vspace{11.0cm}
\includegraphics{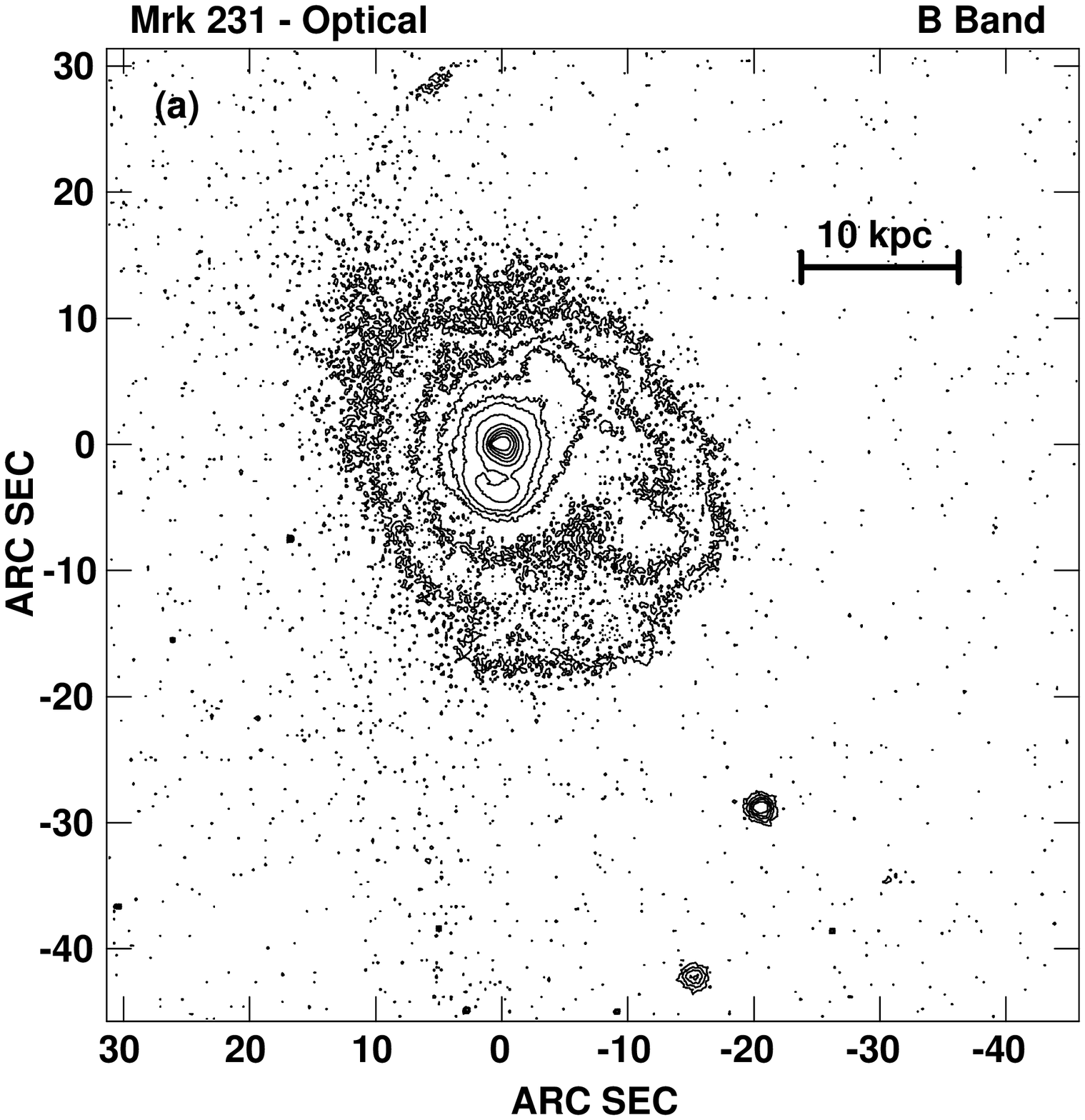}
\includegraphics{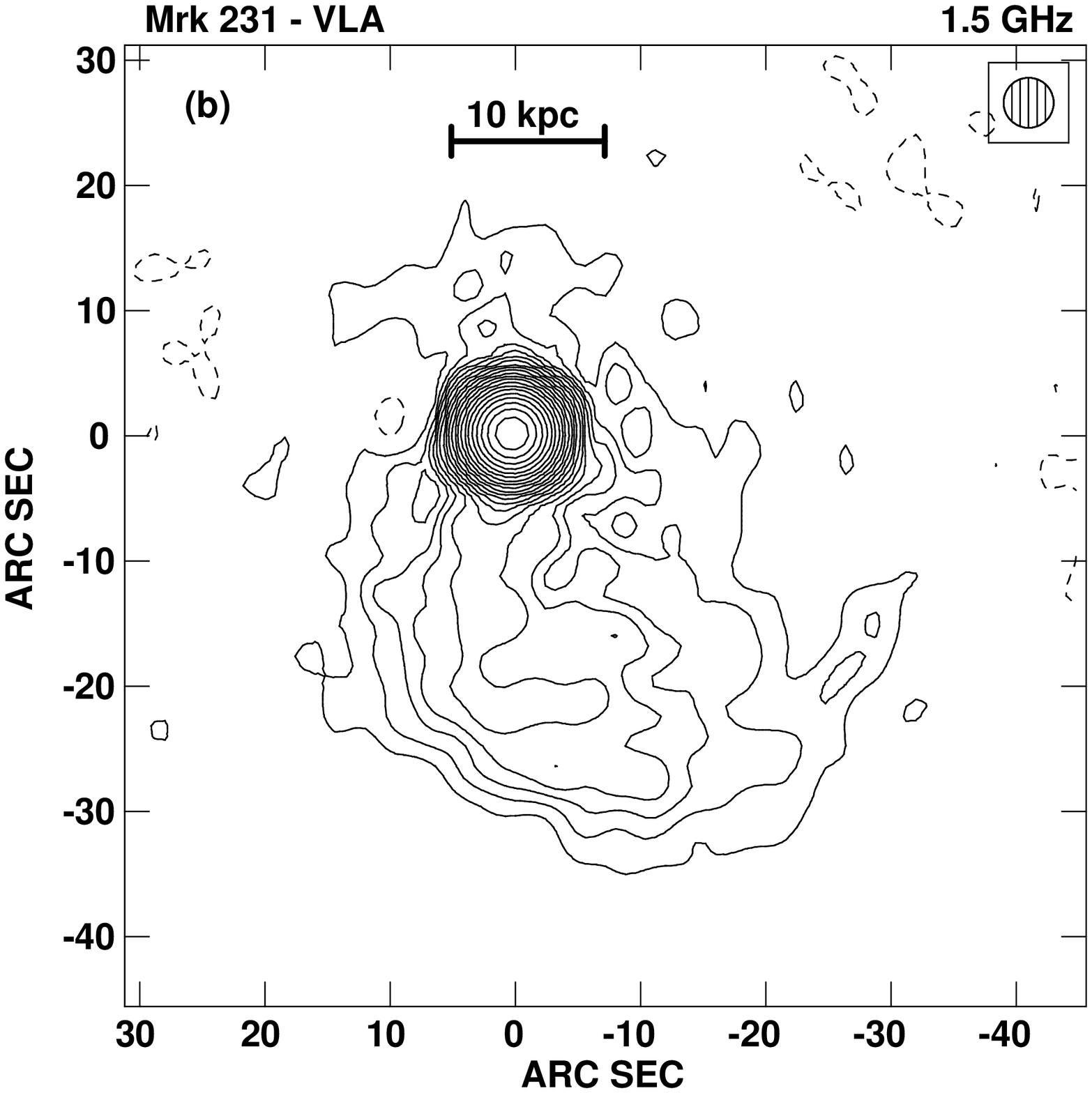}
\includegraphics{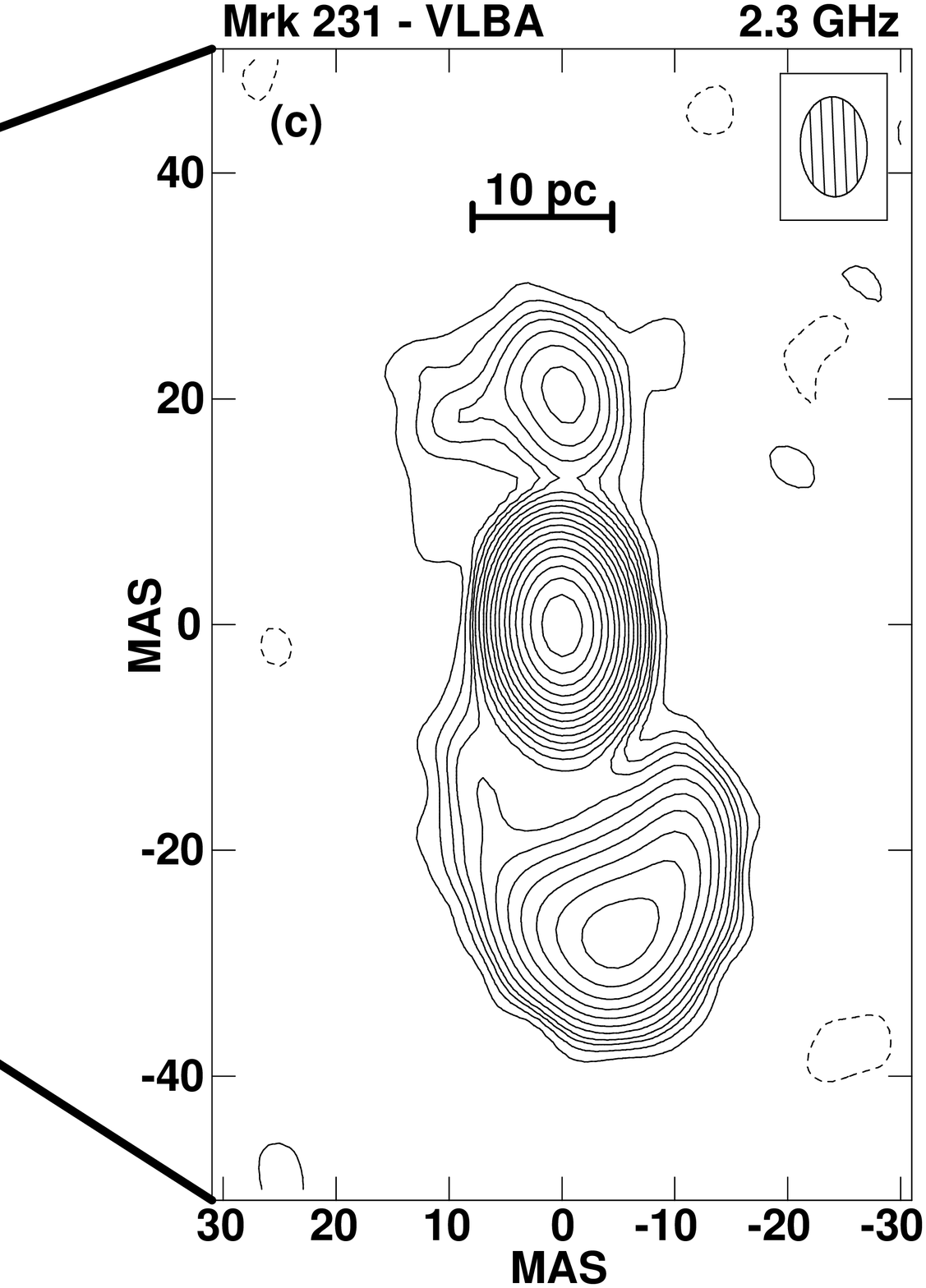}
\includegraphics{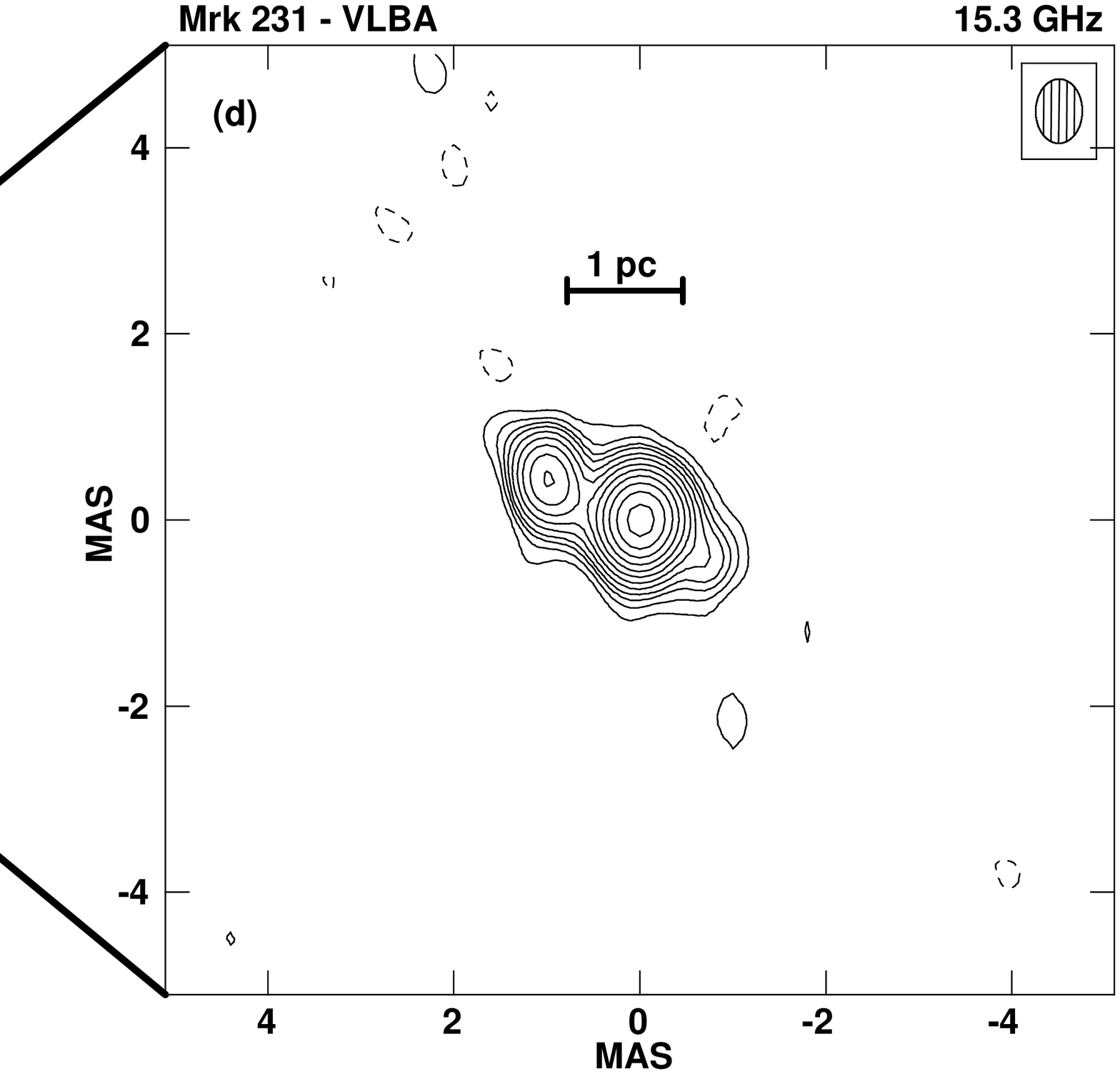}
\caption{Montage of optical and radio images of Mrk~231.
(a) B-band image from Hamilton \& Keel (1987).  Contour intervals are 
factors of $2^{1/2}$ from 0.25\% to 1\% of the peak, and factors
of 2 thereafter to 64\% of the peak.
(b) VLA 1.5-GHz image, at the same scale as (a), with the
two images aligned according to their respective peaks.
Contours are identical to Figure~\ref{fig:ymatch}a.
(c) Full-resolution VLBA 2.3-GHz image of the core seen in the VLA image.
Contours are identical to Figure~\ref{fig:vfull}a.
(d) Magnification of full-resolution VLBA 15-GHz image of the central source seen in 
(c).  Contours are identical to Figure~\ref{fig:vfull}d.
\label{fig:montage}}
\end{figure}

Figure~\ref{fig:montage} summarizes the overall structure of Mrk~231
on a variety of scales.  The two left-hand panels 
(Figures~\ref{fig:montage}a and \ref{fig:montage}b) are an optical
B-band image of the galaxy from Hamilton \& Keel (1987), and, at the same 
scale, the 1.5-GHz VLA image from data taken in 1989.
These panels show that the radio emission
to the south of the nucleus actually extends well beyond the
dominant emission from the optical galaxy.  The core of the
VLA image contains a north-south radio source imaged with the
VLBA (Figure~\ref{fig:montage}c) on very much smaller scales,
with a total extent of $\sim 40$~pc.  Finally, the nucleus of the
galaxy shows additional structures on the 1-pc scale at the 
highest resolution available with the VLBA at 15~GHz, 
as shown in Figure~\ref{fig:montage}d.

\subsection{Kiloparsec Scale}

Mrk~231 is an ultraluminous infrared galaxy with a total luminosity in
excess of $10^{12}~L_\odot$ (\cite{soi89}).  
For over 20 years, it has been known to
contain radio emission about an arcminute in extent (\cite{deb76}),
somewhat larger than the optical galaxy.  The bulk of this emission
comes from a diffuse region within about 30\arcsec\ to the south of
the galaxy nucleus.  Off-nuclear optical imaging and spectroscopy
(\cite{ham87}; \cite{hut87}) revealed H$\alpha$ emission in an
apparent region of star formation centered roughly
10\arcsec--15\arcsec\ to the south of the nucleus.  From the new VLA
imaging at 1.4~GHz, the total flux density in the diffuse emission is
42~mJy.  This emission is primarily concentrated in a
higher brightness feature extending due south of the nucleus for about
20\arcsec\ (16~kpc), then appearing to curve toward the west, as
described previously by Baum et al.\ (1993).  Several different
possibilities for the origin of the diffuse emission are considered
below.

\subsubsection{Thermal Radio Emission from H{\/\sc ii} Regions?}

Most of the diffuse emission south of the nucleus has a
steep spectrum, with a spectral index near $-1.0$, as shown by the
scaled-array observations from 1.5 to 15~GHz.  On spectral grounds
alone, it seems unlikely that thermal processes can make a substantial
contribution.  In addition, it is possible to use the
H$\alpha$ surface brightness measured by Hamilton \& Keel (1987) 
to compute the expected amount of thermal emission from
H{\sc ii} regions.  We estimate the total H$\beta$ flux 
(assuming Case B recombination) in
a 4\arcsec\ VLA beam 12\arcsec\ south of the nucleus to be
$\sim 2\times
10^{-16}$~erg~cm$^{-2}$~s$^{-1}$.  The predicted thermal radio
brightness at 4.9~GHz (e.g., Ulvestad et al. 1981; \cite{con92}) then 
would be only about 1~$\mu$Jy~beam$^{-1}$.  In
contrast, the peak in the diffuse source at 4.9~GHz is $\sim
400$~$\mu$Jy~beam$^{-1}$.  Furthermore, the overall spectrum sets an
upper limit of $\sim 100$~$\mu$Jy~beam$^{-1}$ for the flat-spectrum
contribution at any point, 
entirely consistent with the prediction from the optical spectroscopy.
The lack of discernible thermal radio emission implies that the
intrinsic H$\beta$ flux can be no more than $\sim 100$ times
higher than that observed, implying an upper limit of $A_V\approx 5$
for the extinction in the star-forming region.

The flattest radio spectrum, with $\alpha\approx -0.4$ between 1.5 and
4.9~GHz, occurs near the peak of the diffuse emission in the 4.9-GHz image.  
This is near the peak in the fractional polarization,
and well beyond the region
of significant optical continuum and line emission (\cite{ham87};
\cite{hut87}).  The apparent lack of optical emission from
young stars implies little or no thermal
contribution to the extended radio emission in this area, so the
flattening of the spectrum must have another cause.

\subsubsection{Nonthermal Radio Emission from Supernova Remnants?}

Assuming a total thermal radio flux density of 1--10~$\mu$Jy in the
emission in the southern diffuse lobe, plus a distance of 166~Mpc, the
formulae given by Condon (1992) imply an ionizing flux of $\sim
10^{52}$~photons~s$^{-1}$ and a star formation rate of $\sim 0.03
M_\odot$~yr$^{-1}$ in stars above $5M_\odot$.  This predicts a
nonthermal radio luminosity, from supernova remnants, of $\sim 3\times
10^{19}$~W~Hz$^{-1}$ at 4.9~GHz.  The corresponding radio flux density
at the distance of Mrk~231 would be $\sim 10$~$\mu$Jy.  This
prediction falls three orders of magnitude short of the total flux
density in the southern VLA lobe, and 1.5 orders of magnitude short of
the peak flux density per beam at the location of the putative H{\sc
ii} region.  Alternatively, using the standard radio/infrared relation
for starbursts (e.g., \cite{con92}), the diffuse 1.4-GHz flux
density of 42~mJy would predict infrared emission of 5--10~Jy,
extended over $\sim$30\arcsec, at 60 and 100~$\mu$m.  This is
inconsistent with the small size ($<$1\arcsec--2\arcsec) found for
the near-, mid-, and far-infrared emission (\cite{mil96}; 
\cite{mat87}; \cite{roc93}). Furthermore, although the diffuse
radio emission and the optical galaxy have somewhat similar
shapes, Figure~\ref{fig:montage}
shows that much of the radio emission is actually located beyond
the region of significant B-band emission in the galaxy,
which would not be expected if the radio emission were
associated with a large population of supernova remnants.
Finally, the high fractional polarization implies a
magnetic field that is ordered on a scale much larger than
that expected from a collection of supernova remnants.
Therefore, we rule out the possibility that the diffuse radio 
emission south of the nucleus is generated by star formation
and supernovae.

\subsubsection{Nonthermal Radio Emission Powered by Radio Jet}

The remaining, favored, possibility is that the diffuse emission is
excited by a jet from the galaxy nucleus.  This inference is supported
by an apparent ridge of slightly higher surface brightness emission
connecting back to the nucleus.  Also, the extremely high polarization
at the outer edge of the diffuse lobe implies a well-ordered magnetic
field, rather than the chaotic field expected from a collection of
supernova remnants.  In addition, the polarization vectors indicate
that the magnetic field appears to wrap around the outer edge of the
lobe, as would be natural for emission fed by a nuclear jet.

The total emission of 42~mJy in this southern lobe at 1.4~GHz
corresponds to $\sim 1.4\times 10^{23}$~W~Hz$^{-1}$, and the total
luminosity between 10~MHz and 100~GHz (assuming equal proton and
electron energies, and a spectral index $\alpha=-1.05$) is $\sim
1.8\times 10^{40}$~erg~s$^{-1}$.  If this luminosity arises from
relativistic particles that uniformly fill a lobe 10~kpc in diameter,
then the physical conditions in the lobe can be estimated
(cf. \cite{pac70}).  The minimum-energy magnetic field is $\sim
10$~$\mu$gauss, the total energy in the lobe is $\sim 1.5\times
10^{56}$~erg, and the synchrotron lifetime is $\sim 3\times 10^8$~yr.
The lobe peaks only $\sim 15$~kpc from the nucleus, and therefore
could be
supplied by a jet with an advance speed of only $\sim 50$~km~s$^{-1}$.
This required velocity is much smaller than the speeds of up to
7800~km~s$^{-1}$ seen on parsec scales in the broad-absorption-line
clouds (e.g., Forster et al. 1995).  The north-south VLBA triple implies
the presence of an energy-supplying jet on smaller scales,
although its speed of advance is not presently known.

It is interesting to note that the kiloparsec-scale radio emission 
emerges from the galaxy core in roughly the same direction
as the strong optical emission seen just south of the core, 
in a PA between $165^\circ$ and $170^\circ$
(see Figure~\ref{fig:montage}).  The secondary optical peak
was shown by Surace et al. (1998) to consist of a series of
star-forming knots.  Therefore, if it is correct to assume
that the kiloparsec-scale emission is energized by a radio jet,
that jet would appear to be related to the star formation;
perhaps the jet has compressed thermal material along its path
enough to trigger a burst of star formation.  The
large-scale jet then appears to curve back to the west only beyond
the main extent of the optical galaxy, but in the same general
sense as the tidal tails imaged by Hamilton \& Keel (1987)
and Surace et al. (1998).

\subsection{Sub-Kiloparsec Scales}

CWU discovered a continuum ``halo'' at
1.4~GHz, with a size of $\sim 440$~mas (350~pc) and containing about
130~mJy.  Comparison of Tables~\ref{tab:nsflux} and \ref{tab:vlaflux}
indicates that the amount of flux density missing between VLBA and VLA
scales ranges from $135\pm 14$~mJy at 1.4~GHz to $50\pm 13$~mJy at
15~GHz.  If this emission is dominated by the 350-pc
disk, then the spectral index of that disk appears to be $-0.41\pm
0.12$ between 1.4 and 15~GHz.  Taylor et al. (1998) have
recently imaged weak extended emission on a 1\arcsec\ (800 pc) 
scale, using the VLBA at 0.3 and 0.6~GHz as well as archival
VLA observations at frequencies of 5~GHz and higher.  They suggest
that this emission comes from a weak ``outer disk'' that may
have a spectral turnover near 8~GHz.  It may be that the emission
on 0.5\arcsec--1.0\arcsec\ scales contains both an optically
thin, steep-spectrum component, and some regions that are free-free
absorbed at the lower frequencies.

The sub-kiloparsec ``milli-halo'' seen in NGC~1275 (3C~84) by
Silver, Taylor, \& Vermeulen (1998) has a similar intrinsic strength
and size to the radio disk in Mrk~231.  The milli-halo in NGC~1275 is
reported to have a steeper spectrum, with $\alpha\sim -0.9$.  
Different spectral shapes for NGC~1275 and Mrk~231 may imply different
emission processes, different electron energy distributions, or just
varying amounts of free-free
absorption.  The suggested model for NGC~1275 is that
the milli-halo is caused by particles leaking out of the radio jet
into the surrounding medium (\cite{sil98}).  However, for Mrk~231, the
elongation of the 350-pc emission perpendicular to the 40-pc VLBA
triple and parallel to the slightly larger scale CO~disk indicates,
instead, a possible relationship to the disk of material thought to
surround active galactic nuclei (e.g., \cite{ant93}).  CWU associate
this elongated emission with a disk containing atomic, molecular,
and dust components on a scale of a few hundred parsecs.  The short
electron lifetimes ($<10^5$~yr) in this disk imply local particle 
acceleration, requiring star formation or other shock processes
in the inner kiloparsec.  In fact, the radio
continuum emission from this disk or torus is likely to be powered by massive
star formation, since it is consistent with the canonical radio-infrared
relation.  This is in agreement with the conclusion of Downes \&
Solomon (1998), who suggest that the bulk of the far-infrared
emission in Mrk~231 is powered by a starburst.  The slight
flattening of the radio spectrum relative to the canonical
starburst spectral index of $\sim -0.8$ could be accounted
for by localized free-free absorption within the disk.

\subsection{10--100 Parsec Scales}

The north-south triple source in Mrk~231 has a total extent of roughly
50~mas, or 40~pc.  At 1.4 and 2.3~GHz, there is considerable emission
detected from the central core out to the outer edges of the triple.
However, at 5.0~GHz, most of the detected extended emission is in the outermost
parts of the lobes.  The high resolution images at 2.3 and 5.0~GHz
(Figure~\ref{fig:vfull}) show that the ends of the lobes are
significantly resolved perpendicular to the direction to the core,
with transverse extents of 15--20~mas ($\sim$12--16~pc).  This may
indicate the presence of shocks at ``working surfaces'' where jets are
attempting to burrow out of the nuclear regions.  The outer edges of
the VLBA triple source are at a distance similar to the inner scale of
the diffuse radio emitter and the H{\sc i}-absorption cloud imaged by
CWU, so it seems plausible that these
lobes are generated by a jet running into the inner surface of an
H{\sc i} shell or disk.  It is also interesting to note that the major
axis of the triple source is similar to the axis of the diffuse radio
source and the elongation of the optical galaxy on scales 2--3 
orders of magnitude larger (see
Figure~\ref{fig:montage}).  This suggests a long-term memory of the
symmetry axis of this Seyfert galaxy, possibly associated with an
accretion disk in its center.

\subsubsection{Northern Component of Triple}

At the resolution of the 1.4-GHz VLBA image, the northern component of
the 40-pc triple has a steep spectrum between 1.4 and 8.4~GHz, with
a spectral index of $-0.99\pm 0.29$.  Within the errors, the
spectrum is straight across the entire flux range, consistent
with optically thin synchrotron emission from an
ensemble of electrons with a steep power-law distribution in 
energy.  However, 
the peak of the northern component shifts systematically toward the
southwest with increasing frequency (see Table~\ref{tab:nsflux}),
implying the presence of spectral gradients within the
lobe.  At 1.4~GHz, this peak is located 23.3~mas (17~pc) from the
core, in PA 6.5\arcdeg.  At the same low resolution, the 8.4-GHz peak
is located only 19.1~mas (14~pc) from the core in PA $-4.3$\arcdeg.

\begin{figure}[th]
\vspace{15.5cm}
\includegraphics{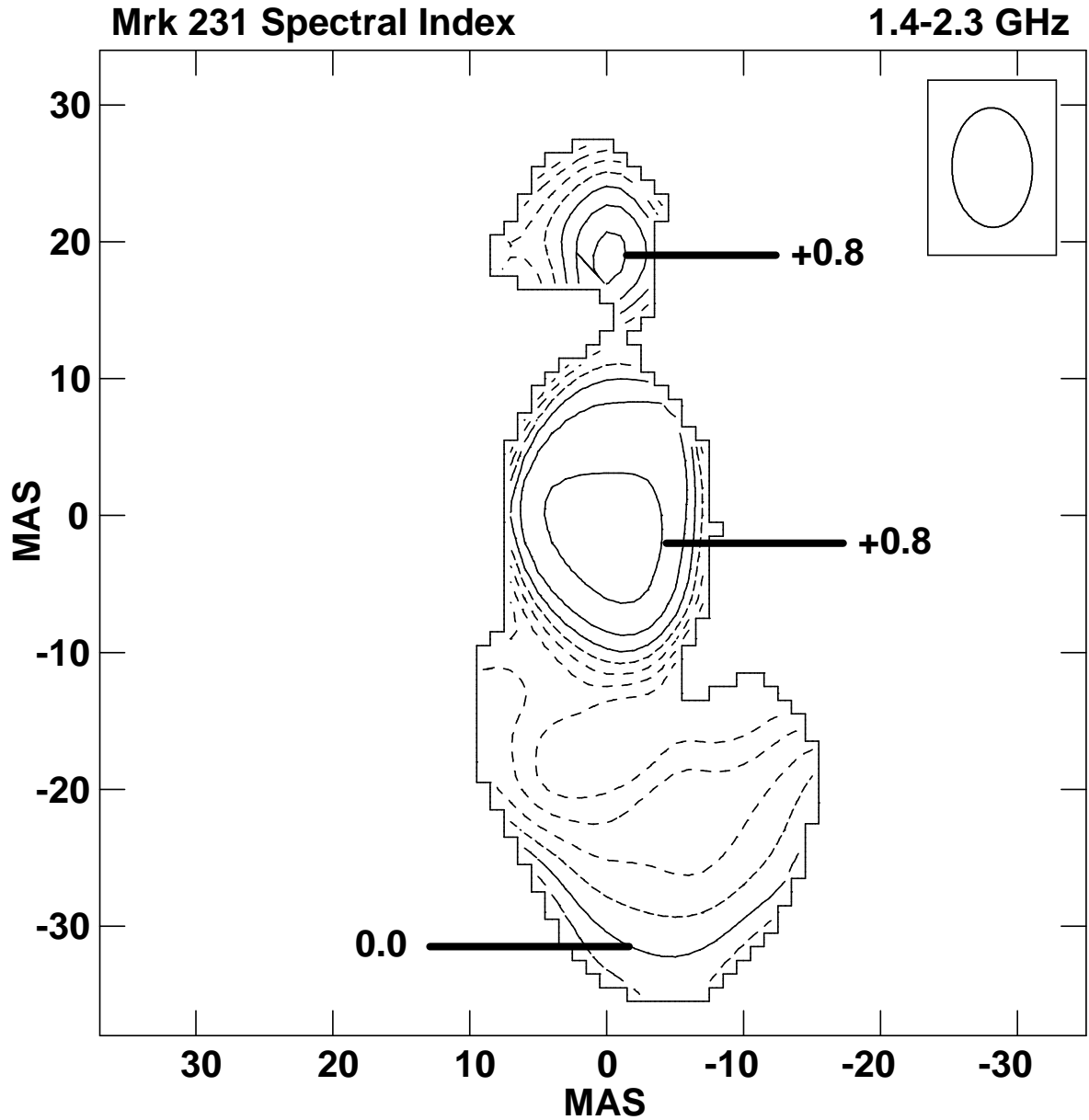}
\caption{Image of the spectral index between 1.4 and 2.3~GHz
in the north-south VLBA triple in Mrk~231, with the input 
1.4- and 2.3-GHz images
blanked at 8 times their respective r.m.s. noise.  Contours are shown at
intervals of 0.4 from $-1.2$ to 1.2, and negative contours are
shown dashed.  A few contours are identified on the image
for reference.
\label{fig:spectral}}
\end{figure}

 Since most of the shift in the peak of the northern lobe 
occurs between 1.4 and 2.3~GHz, a spectral index image was made
between those two frequencies, using the 2.3-GHz image tapered to the
1.4-GHz resolution.  That image, displayed in Figure~\ref{fig:spectral},
shows the
presence of a region with an inverted spectrum, roughly 18~mas
(14~pc) north of the galaxy nucleus, although the point-spread
function at 1.4-GHz is too large to cleanly resolve the region.
(The sharp edges in the spectral-index map are caused by 
blanking the individual input maps at 8 times their respective
noise levels.) A similar image of the spectral index 
between 2.3 and 5~GHz, at the 2.3-GHz resolution, shows
a small region with a nearly flat spectrum between those two
frequencies near the same location.  This implies that the northern
lobe may contain a small component, with a diameter no larger
than $\sim 4$~mas (3~pc), whose spectrum turns over near a
frequency of 2--3~GHz.

The most logical cause for the apparent spectral turnover in the
northern lobe is free-free absorption in an ionized region with a
temperature near $10^4$~K.  For a turnover frequency of 2~GHz, the
emission measure would be $1.3\times 10^7$~cm$^{-6}$~pc.
Both the large variation in spectral index and the shift of the peak
as a function of frequency occur over 3--4~mas, implying that the
absorbing medium has a size of $\sim 3$~pc or less on the plane of the
sky.  If this dimension is also used as an estimate of the
line-of-sight distance through the absorber, then the average density
along a 3-pc line of sight would be $\sim 2\times 10^3$~cm$^{-3}$.
Such a density is, for instance, fairly typical for ionized clouds in
Seyfert narrow-line regions (e.g., \cite{ost93}).

\subsubsection{Southern Component of Triple}

The spectral-index image also indicates a spectral gradient in the
southern VLBA component between 1.4 and 2.3~GHz, including a portion
at the outer edge of the lobe with an inverted spectrum.  The spectral
index image between 2.3 and 5.0~GHz shows no such inverted component.
The presence of an inverted component between 1.4 and 2.3~GHz is
further confirmed by the overall lobe spectrum, which has 
$\alpha = -0.41\pm 0.29$ between 1.4 and 2.3~GHz and
$\alpha = -1.54\pm 0.20$ between 2.3 and 8.4~GHz 
(cf. Table~\ref{tab:nsflux}
and Figure~\ref{fig:vspectra}).  If the steep spectrum
above 2.3~GHz continued to 1.4~GHz, the flux density at that
frequency would be 65~mJy instead of the measured value of 37~mJy,
implying that a substantial fraction of the total flux of the lobe is
absorbed at the lowest frequency.  Hypothesizing a turnover frequency
due to free-free absorption midway between our two lowest observing
frequencies, near 1.8~GHz, the emission measure would be $1\times
10^7$~cm$^{-6}$pc.  The region of strongest absorption appears to be
at the outer edge of the lobe, between 22 and 28~mas ($\sim 20$~pc)
from the central
component.  This implies an absorbing cloud with a size near 5~pc, and
the derived particle density is then $\sim 1.5\times 10^3$~cm$^{-3}$.
Both the density and the size of this ionized cloud are consistent
with those required for free-free absorption in the northern
component.

The overall high-frequency spectrum could be artificially steepened by
resolution effects, which would tend to decrease the flux density at
the higher frequencies.  However, the spectral index between 2.3 and
5~GHz is still near $-1.5$ in images made at the full 2.3-GHz
resolution, and the mix of short and long spacings on the VLBA
implies that little 5-GHz emission should be missing on this
scale.  Therefore, the overall spectrum of the southern VLBA lobe
is indeed quite steep intrinsically.

\subsubsection{Overall Structure of North-South VLBA Triple}

The 40-pc VLBA triple appears to be significantly affected by
free-free absorption due to ionized gas.  On a somewhat larger scale,
a neutral absorbing component has been detected in our H{\sc i} absorption
study (\cite{car98}).  Therefore, a possible inference is that the
ionized gas is merely the inner part of the putative disk seen in H{\sc
i}, ionized by the central continuum source in Mrk~231.  That disk may
be a larger version of the disks traced by H$_2$O maser emission in 
active galaxies such as NGC~4258 (\cite{miy95}; \cite{her96}; 
\cite{her97}).  However, the viewing angle for the inner disk in Mrk~231
(e.g., \cite{car98}) would be such that no H$_2$O maser emission is present.
The much larger size of the apparent disk in Mrk~231 could be related to 
its much more luminous central continuum source.

The north-south VLBA source in Mrk~231 bears a striking resemblance to
that seen in 3C~84 (\cite{ver94}; \cite{wal94}; \cite{dha98}).
NGC~1275, the host galaxy of 3C~84, is slightly more than twice as
close to us as Mrk~231, and the size of its north-south
radio source is 10~pc, about a quarter the size of the triple in
Mrk~231.  In addition, the northern component of 3C~84 (the
``counterjet'') also exhibits free-free absorption near 15~GHz,
attributed to ionized gas with a path length of several parsecs and a
density above $10^4$~cm$^{-3}$ (Vermeulen et al. 1994).  Models of the
absorption in 3C~84 indicate that the absorbing gas is not 
spherically distributed, but can be in a warped disk that could be
kept ionized by a central continuum source (\cite{lev95}).  However,
in 3C~84, the disk appears to block only the northern VLBA component,
whereas in Mrk~231 the apparent free-free absorption toward parts of both 
the southern and northern components indicate
that the disk may block both components.  This blockage occurs despite
the fact that the VLBA lobes in Mrk~231 are several times farther from
the radio core than is the absorbed region of the jet in 3C~84.
Levinson et al.\ (1995) estimated a total bolometric luminosity of
$\sim 2\times 10^{11}~L_\odot$ (scaling their value to $H_0=75$) for
3C~84, while the total luminosity of Mrk~231 is $\sim 15$ times
greater.  This fact, combined with the inference that the ionized
density in Mrk~231 is $\sim 10$ times smaller than that derived from
the higher-frequency spectral turnover in 3C~84 (Vermeulen et al. 1994),
indicates that the volume of the ionized region could be
hundreds of times larger in Mrk~231 than in 3C~84.  This may account
for the fact that free-free absorption is detected over a much larger
scale in Mrk~231 than in 3C~84.

An alternative explanation for the free-free absorption is connected
with the possibility that much of the infrared emission in Mrk~231
may be generated by star formation rather than by an active
galactic nucleus (\cite{dow98}).  In this case, a local source
of ionizing radiation could account for the free-free-absorbing
clouds.  Ionization of spherical clouds 5~pc in diameter, with
densities of $10^3$~cm$^{-3}$, requires only $\sim 10^{50}$
ionizing photons per second.  This ionization rate could be
accounted for by just
a few early O stars (\cite{ost89}), easily consistent with 
the high infrared luminosity.

\subsection{Parsec Scales}

The central source of the 40-pc VLBA triple undoubtedly contains the
actual nucleus of Mrk~231.  This central source was shown in
Section~\ref{central} (see Table~\ref{tab:cflux}
and Figure~\ref{fig:vfull}) to consist of two dominant components
separated by 1.1~mas (0.9~pc) in PA $\sim 65$\arcdeg, with a possible
weak third component located 0.85~mas (0.7~pc) to the southwest.  At
8.4~GHz, the two stronger components have brightness temperatures of
$9\times 10^9$~K and $>1\times 10^9$~K.  The bulk of the flux density
in the strongest component is unresolved, so it may have a peak
brightness temperature considerably exceeding $10^{10}$~K.

The spectrum of component C peaks between 5.0 and 8.4~GHz (see
Figure~\ref{fig:vspectra}); the spectral index between 8.4 and 15~GHz
is $-1.32\pm 0.23$, indicative of optically thin synchrotron
radiation.  Table~\ref{tab:cflux} indicates that the two dominant
components of this source both have steep spectra between 8.4 and
15~GHz. The high-frequency (``intrinsic,'' or un-absorbed) spectrum of
the stronger component may be steeper than that of the weaker
component, but this is somewhat uncertain due to the blending of the
two components at 8.4~GHz.

Two possibilities for the spectral turnover are either synchrotron
self-absorption or free-free absorption similar to that deduced for
the northern and southern lobes.  The stronger component has an
8.4-GHz brightness temperature of $\sim 10^{10}$~K, indicating that
synchrotron self-absorption is a possible cause for the spectral
turnover.  If the turnover occurs near 6~GHz at a flux density near
150~mJy, the 8.4-GHz size of $0.56 \times 0.40$~mas implies
a magnetic field strength near 0.5~gauss.  For comparison, the
minimum-energy magnetic field calculated for a straight spectrum with
a spectral index of $-1.3$ between $10^7$ and $10^{11}$~Hz, and with
equal proton and electron energy densities, is similar, $\sim 0.2$~gauss.  
The equipartition magnetic field is thus near the value required for
synchrotron-self-absorption to occur in the stronger component, making
it likely that this could account for the
overall turnover of the spectrum.  Alternatively, free-free absorption
causing a turnover near 6~GHz,
 and over a 1-pc path length, would require an ionized
density of $\sim 1.1\times 10^4$~cm$^{-3}$.

Attributing the turnover above 5~GHz to synchrotron self-absorption in
the stronger central component does not eliminate the possibility that
free-free absorption also occurs at a somewhat lower frequency.  
For example, the
northeastern component of the central source has a spectral index of
$\sim -0.7$ between 8.4 and 15~GHz, as indicated by the two-component
fit to the central source.  An extrapolation of this spectrum to lower
frequencies would predict a flux density of $\sim$45~mJy at 2.3~GHz
and $\sim$65~mJy at 1.4~GHz.  The latter prediction exceeds the total
observed value of 53~mJy for all of component C.  This
contradiction can be removed by postulating the presence of either 
free-free or synchrotron
absorption of the weaker component in the vicinity of 2~GHz.  Such
absorption is consistent with the results for the outer VLBA lobes,
again indicating the presence of a large quantity of ionized gas in
the inner regions of Mrk~231.  However, strong variability of the
northeastern component between 1996 and 1998 (Ulvestad et al.,
1998b, and in preparation) implies that it may well be the location of the
actual nucleus; its brightness temperature may be considerably
greater than the lower limit of $10^9$~K found in this work,
consistent with the possible presence of synchrotron self-absorption.

The major axis of the central VLBA source is near 65\arcdeg, very
different from the position angle of about 0\arcdeg--5\arcdeg\ for the
larger-scale VLBA triple.  If we assume (based on our recent
observations of variability) that the northeastern component of
the central source is the galaxy core, and the inner radio jet
extends to the southwest in position angle 115\arcdeg, the obvious
hypothesis would be that this jet must twist within the inner few
parsecs of the galaxy to feed the more distant VLBA lobes.  However,
there is no evidence in any VLBA images for a direct connection 
between the 1-pc-scale source and the larger scale
VLBI lobes.  Apparent large changes in VLBA position angles
within the inner few parsecs are also present in 
NGC~4151 (\cite{ulv98a}), 
so this morphological trait is not unique to Mrk~231.
Indeed, a similar circumstance was discovered recently in the nearest
active galaxy, Centaurus~A, where the sub-parsec axis defined by the
VLBI radio jet (\cite{jon96}; \cite{tin98}) is misaligned by some
70\arcdeg\ from the axis of the infrared disk imaged with HST
on a 40-pc scale (\cite{sch98}).  An alternative hypothesis
to jet curvature would be 
that the northeastern component of the central source is part of an
accretion disk or torus as claimed for component S1 in NGC~1068
(\cite{gal97}).  However, this seems untenable for Mrk~231; the brightness
temperature in Mrk~231 is more than 1000 times higher than that in
NGC~1068, so an interpretation as thermal bremsstrahlung or reflected
synchrotron emission, while reasonable for NGC~1068, becomes
implausible for Mrk~231.

If the central source in Mrk~231 is the inner portion of a
jet, its symmetry axis would presumably represent the inner axis of the
accretion disk around the black hole that powers the active galactic
nucleus.  In contrast, the VLBA triple, plus the H{\sc i} and CO
disks, clearly indicate a very different axis on scales greater than a
few parsecs.  The smaller scale is typical for the optical
broad-line region, while the larger scale matches that for the optical
narrow-line region in Seyferts.  Therefore, a possible interpretation is
that the broad-line and narrow-line regions have symmetry axes
that are related only over the long term.  
The simplest unified schemes for Seyfert galaxies
(\cite{ant93}) suggest that Seyfert~1 galaxies like Mrk~231 are those
in which the line of sight to the broad-line region misses the nuclear
disk or torus, while Seyfert~2 galaxies are those in which our viewing
angle prevents a direct view of the broad-line region.  These models
typically rely on a common symmetry axis for the broad-line and
narrow-line regions, whereas the data presented here indicate that
those axes may be very different, at least at the current epoch.  
This is in accord with recent
calculations (Pringle 1996, 1997; \cite{mal96}), 
showing that the accretion disks
in active galactic nuclei may be severely warped by the local
radiation source, and also with observations of 
sub-parsec-scale warped disks traced out by VLBI imaging of 
water megamasers in NGC~4258 (Herrnstein et al. 1996).
One could imagine that the larger scale VLBA and
VLA sources represent the ``average'' axis of the disk, 
while the small-scale VLBA source shows the instantaneous direction of 
an inner disk that might be determined by the gas most recently
added to that disk.  The amount of jet curvature, or disk precession
or warping, required 
would be reduced considerably if we happen to be viewing the radio
jet nearly end-on, so that a small change in angle could 
appear much larger in projection.

\section{Summary}

The VLBA and VLA have been used to image the continuum emission from
Mrk~231 on scales ranging from parsecs to kiloparsecs.  
An asymmetric, diffuse radio source is traced for more than 25~kpc.
It exhibits linear polarization as high as 57\%, has a ridge of
modest brightness aligned with a starburst region several kiloparsecs
south of the galaxy nucleus and (roughly) with the 40-pc VLBA triple source, and
appears to be powered by energy deposition from the jet.
This diffuse radio source extends beyond the bulk of the optical
emission from Mrk~231.  Inside the diffuse radio emission,
a 350-pc disk of radio continuum and H{\sc i} (CWU),
appears to be caused by synchrotron emission
that may be free-free absorbed in some regions.
This is consistent with the inference
that the disk is powered by a massive starburst.

The 40-pc VLBA triple source exhibits free-free absorption in both its
northern and southern components at frequencies of 2--3~GHz, implying
the presence of ionized gas clouds several parsecs in diameter with
densities of 1--2$\times 10^3$~cm$^{-3}$.  These clouds may reside in
the ionized inner region of the same disk responsible for the H{\sc i}
absorption (\cite{car98}).  Their ionization may be powered either by
the active galactic nucleus or by a few O stars.
The central component of the 40-pc 
source has an elongation PA $\sim 65$\arcdeg\
from that of the 40-pc VLBA triple.  This core contains at least two
components, with brightness temperatures of $\sim 10^{10}$~K
and $> 10^9$~K, each of which is
absorbed at low frequencies.  The data are consistent with the
presence of synchrotron self-absorption between 5 and 10~GHz in
the stronger component, implying a
magnetic field of $\sim 0.5$~gauss.  The weaker component,
about 1~pc to the northeast, is absorbed below 2--3~GHz, due
either to free-free absorption or synchrotron self-absorption.

All data support the presence of two symmetry axes in Mrk~231:
(1) an inner axis in PA $\sim 65$\arcdeg\ that determines the initial
direction of the radio jet, and is likely to be associated with the
current axis of a central black hole or its accretion disk; 
and (2) an outer axis that is near PA
0\arcdeg, perpendicular to a disk or torus that is ionized in its inner
10--20~pc radius, neutral out to $\sim 200$~pc from the center, and
then molecular out to $\gtrsim 400$~pc radius.  The outer axis presumably
represents the average long-term axis of the central engine.
These two axes are substantially misaligned unless the parsec-scale
radio source is viewed nearly end-on, so that small curvature
would be enhanced greatly by projection.

\acknowledgments

We are grateful to Bill Keel for supplying the
optical image of Mrk~231, and to Greg Taylor for useful
discussions.  This research has made use of the NASA/IPAC
Extragalactic Database (NED) which is operated by the Jet Propulsion
Laboratory, California Institute of Technology, under contract with
the National Aeronautics and Space Administration.

\clearpage

\begin{deluxetable}{lclrccc}
\tablecolumns{7}
\tablewidth{0pc}
\tablecaption{Log of VLBA Observations of Mrk~231}
\tablehead{
                       &
\colhead{UT}           & 
                       &
                       &
                       &
                       &
\colhead{Integration}  \\
\colhead{Epoch}        & 
\colhead{Range}        & 
\colhead{Telescopes}   & 
\colhead{Frequency}    & 
\colhead{Bandwidth}    & 
\colhead{Polarization} &
\colhead{Time}         \\
&\colhead{(hr)}&& \colhead{(GHz)} & \colhead{(MHz)} & & \colhead{(min)} }
\startdata
1995 Nov 13\ldots & 11--23 & VLBA    &  1.367~~ & 32 & LCP  & 154  \nl
                  &       &         &  2.271~~ & 32 & RCP  & 143  \nl
                  &       &         &  4.987~~ & 32 & LCP  & 132  \nl
1996 Dec 8\ldots  & 09--21 & VLBA+Y1\tablenotemark{a} &  4.987~~ & 32 & LCP  & 152 \nl
                  &       &         &  8.421~~ & 32 & RCP  & 152 \nl
                  &       &         & 15.365~~ & 32 & LCP  & 156 \nl
1996 Dec 27\ldots & 08--20 & VLBA+YP\tablenotemark{b} &  
1.367\tablenotemark{c} ~~ & 16 & Dual & 380 \nl
                  &       &         & 22.233~~ & 32 & LCP  &  96 \nl
\enddata
\tablenotetext{a}{Y1 denotes a single 25-m antenna at the VLA.}
\tablenotetext{b}{YP denotes the phased VLA. Twenty-five 25-m 
antennas were available.}
\tablenotetext{c}{Data are from the H{\sc i} line study by 
CWU.}
\label{tab:vlbaobs}
\end{deluxetable}
\clearpage

\begin{deluxetable}{lcrcr}
\tablecolumns{5}
\tablewidth{0pc}
\tablecaption{Log of VLA Observations of Mrk~231}
\tablehead{
                        & 
                        & 
                        &
                        & 
\colhead{Integration}   \\
\colhead{Epoch}         & 
\colhead{Configuration} & 
\colhead{Frequency}     & 
\colhead{Bandwidth}     & 
\colhead{Time}          \\
& & \colhead{(GHz)} & \colhead{(MHz)} & \colhead{(min)} }
\startdata
1988 Aug 10\ldots & D & 14.940~~ & 50 & 186~~~~~ \nl
1989 Apr 14\ldots & B &  1.490~~ & 50 &  86~~~~~ \nl
1989 Sep 1\ldots  & C &  4.860~~ & 50 &  88~~~~~ \nl
1995 Nov 17\ldots & B &  1.425~~ & 50 &  11~~~~~ \nl
                  &   &  4.860~~ & 50 &  11~~~~~ \nl
                  &   &  8.440~~ & 50 &  11~~~~~ \nl
                  &   & 14.940~~ & 50 &  11~~~~~ \nl
                  &   & 22.460~~ & 50 &  11~~~~~ \nl
1996 Dec 27\ldots   & A &  1.365~~ & 25 & 380~~~~~ \nl
                  &   &  4.985~~ & 50 &   2~~~~~ \nl
                  &   &  8.415~~ & 50 &   2~~~~~ \nl
                  &   & 15.365~~ & 50 &   2~~~~~ \nl
                  &   & 22.235~~ & 50 &  96~~~~~ \nl
\enddata
\label{tab:vlaobs}
\end{deluxetable}
\clearpage

\begin{deluxetable}{lrcrr}
\tablecolumns{5}
\tablewidth{0pc}
\tablecaption{Properties of VLBA North-South Triple at 1.4-GHz Resolution}
\tablehead{
\colhead{Component} & & \colhead{Total} & \colhead{R.A.} & \colhead{Decl.} \\
\colhead{Label}        & 
\colhead{Frequency}    & 
\colhead{Flux Density} & 
\colhead{Offset}       & 
\colhead{Offset}       \\
& \colhead{(GHz)} & \colhead{(mJy)} & \colhead{(mas)} & \colhead{(mas)} }
\startdata
N\ldots &  1.4~~~ & $  6\pm 1  $ & 2.64~~    &    23.19~~ \nl
        &  2.3~~~ & $  4\pm 1  $ & $-$0.17~~ &    20.56~~ \nl
        &  5.0~~~ & $  2\pm 1  $ & $-$0.58~~ &    19.63~~ \nl
        &  8.4~~~ & $  1\pm 0.5$ & $-$1.45~~ &    19.07~~ \nl
C\ldots &  1.4~~~ & $ 53\pm 5  $ & 0.00~~    &     0.00~~ \nl
        &  2.3~~~ & $ 97\pm 10 $ & 0.00~~    &     0.00~~ \nl
        &  5.0~~~ & $163\pm 16 $ & 0.00~~    &     0.00~~ \nl
        &  8.4~~~ & $137\pm 14 $ & 0.00~~    &     0.00~~ \nl
        & 15.4~~~ & $ 62\pm 6  $ & 0.00~~    &     0.00~~ \nl
S\ldots &  1.4~~~ & $ 37\pm 4  $ & $-$5.34~~ & $-$26.50~~ \nl 
        &  2.3~~~ & $ 30\pm 3  $ & $-$5.06~~ & $-$27.48~~ \nl
        &  5.0~~~ & $  8\pm 2  $ & $-$4.42~~ & $-$28.28~~ \nl
        &  8.4~~~ & $  4\pm 1  $ & $-$5.65~~ & $-$27.67~~ \nl
\enddata
\label{tab:nsflux}
\end{deluxetable}
\clearpage

\begin{deluxetable}{lrccrr}
\tablecolumns{6}
\tablewidth{0pc}
\tablecaption{VLBA Model Fits to Component C at Matched Resolution}
\tablehead{
\colhead{Component} & & \colhead{Total} & & 
\colhead{R.A.} & \colhead{Decl.} \\
\colhead{Label}        & 
\colhead{Frequency}    & 
\colhead{Flux Density} & 
\colhead{Size}         &
\colhead{Offset}       & 
\colhead{Offset}       \\
& \colhead{(GHz)} & \colhead{(mJy)} & \colhead{(mas)} & \colhead{(mas)} & 
\colhead{(mas)} }
\startdata
\multicolumn{6}{c}{\underbar{Single-Component Fits at 5-GHz Resolution}} \nl
C\ldots &  5.0~~~ & $162\pm 16 $ & $0.82\times 0.43$, PA 63.0\arcdeg &
  0.00 & 0.00 \nl 
 & 8.4~~~ & $134\pm 13$& $1.05 \times 0.48$, PA 61.8\arcdeg &
  0.00 & 0.00 \nl
 &15.4~~ & $62\pm 6$& \nodata & 0.00 & 0.00 \nl
\multicolumn{6}{c}{\underbar{Two-Component Fits at 8.4-GHz Resolution}} \nl
C1\ldots &8.4~~~ & $114\pm 11$& $0.56\times 0.40$, PA 70.1\arcdeg & 0.00 &
0.00 \nl
C2\ldots &8.4~~~ & $18\pm 2$& \nodata & 1.00 & 0.49 \nl
C1\ldots & 15.4~~ & $51\pm 5$& $0.52\times 0.24$, PA 70.8\arcdeg & 0.00 & 
0.00  \nl
C2\ldots & 15.4~~ & $12\pm 2$& \nodata & 1.03 & 0.44 \nl
\multicolumn{6}{c}{\underbar{Three-Component Fit at 15-GHz Resolution}} \nl
C1\ldots & 15.4~~ & $49\pm 5$ & $0.37 \times 0.24$, PA 92.0\arcdeg & 0.00 &
0.00 \nl
C2\ldots & 15.4~~ & $13\pm 2$ & \nodata & 0.98 & 0.43 \nl
C3\ldots & 15.4~~ & $3\pm 1$ & \nodata & $-$0.74 & $-$0.42 \nl
\enddata
\label{tab:cflux}
\end{deluxetable}
\clearpage

\begin{deluxetable}{rccc}
\tablecolumns{4}
\tablewidth{0pc}
\tablecaption{VLA Results at Matched Resolution, 1988--1989}
\tablehead{
& \colhead{Core} & \colhead{Extended} & \colhead{Extended} \\
\colhead{Frequency}      & 
\colhead{Flux Density}   &
\colhead{Flux Density}   &
\colhead{Peak Intensity} \\
\colhead{(GHz)} & \colhead{(mJy)} & 
\colhead{(mJy)} & \colhead{(mJy beam$^{-1}$)} }
\startdata
 1.5~~~ & $238\pm 12$ & $42\pm 4 $ & 1.3     \\
 4.9~~~ & $343\pm 17$ & $12\pm 2 $ & 0.5     \\
15.0~~~ & $279\pm 14$ & $< 5     $ & $< 0.3$ \\
\enddata
\label{tab:vlascale}
\end{deluxetable}
\clearpage

\begin{deluxetable}{rcrc}
\tablecolumns{4}
\tablewidth{0pc}
\tablecaption{VLA Core Flux Densities}
\tablehead{
\multicolumn{2}{c}{1995 November} & 
\multicolumn{2}{c}{1996 December} \\
& \colhead{Core} & & \colhead{Core} \\
\colhead{Frequency}    & 
\colhead{Flux Density} &
\colhead{Frequency}    &
\colhead{Flux Density} \\
\colhead{(GHz)} & \colhead{(mJy)} & 
\colhead{(GHz)} & \colhead{(mJy)} } 
\startdata
 1.425~~ & $241\pm 12$ &  1.365~~ & $231\pm 12$ \\
 4.860~~ & $254\pm 13$ &  4.985~~ & $270\pm 13$ \\
 8.440~~ & $179\pm 9 $ &  8.415~~ & $203\pm 15$ \\
14.940~~ & $135\pm 7 $ & 15.365~~ & $112\pm 11$ \\
22.460~~ & $136\pm 7 $ & 22.235~~ & $ 62\pm 9 $ \\
\enddata
\label{tab:vlaflux}
\end{deluxetable}
\clearpage

\clearpage

\clearpage

\clearpage

\clearpage

\clearpage

\clearpage

\end{document}